\documentclass{emulateapj}
\providecommand{\comment}[1]{{\bf(#1)}}
\renewcommand{\comment}[1]{}
\usepackage{graphicx}
\usepackage{natbib}
\usepackage{aas_macros}
\usepackage{amsmath}
\usepackage{amssymb}
\usepackage{fancyvrb}
\usepackage{epsf}
\usepackage{sidecap}

\usepackage{grffile}

\shorttitle{PRIMUS: Galaxy Clustering as a Function of Luminosity and Color}
\shortauthors{Skibba, Smith, Coil, et al.}

\begin{document}

\title{PRIMUS: Galaxy Clustering as a Function of Luminosity and Color at $0.2<$\lowercase{\textit{z}}$<1$}

\author{
Ramin~A.~Skibba\altaffilmark{1},
M.~Stephen~M.~Smith\altaffilmark{1},
Alison~L.~Coil\altaffilmark{1,2},
John~Moustakas\altaffilmark{3},
James~Aird\altaffilmark{4},
Michael~R.~Blanton\altaffilmark{5},
Aaron~D.~Bray\altaffilmark{6}, 
Richard~J.~Cool\altaffilmark{7},
Daniel~J.~Eisenstein\altaffilmark{6},
Alexander~J.~Mendez\altaffilmark{1},
Kenneth~C.~Wong\altaffilmark{8},
Guangtun~Zhu\altaffilmark{9} 
}
\altaffiltext{1}{Department of Physics, Center for Astrophysics and Space Sciences, University of California, 9500 Gilman Dr., La Jolla, San Diego, CA 92093; rskibba@ucsd.edu}
\altaffiltext{2}{Alfred P. Sloan Foundation Fellow}
\altaffiltext{3}{Department of Physics and Astronomy, Siena College, 515 Loudon Road, Loudonville, NY 12211, USA}
\altaffiltext{4}{Department of Physics, Durham University, Durham DH1 3LE, UK}
\altaffiltext{5}{Center for Cosmology and Particle Physics, Department of Physics, New York University, 4 Washington Place, New York, NY 10003}
\altaffiltext{6}{Harvard-Smithsonian Center for Astrophysics, 60 Garden Street, Cambridge, MA 02138, USA}
\altaffiltext{7}{MMT Observatory, 1540 E Second Street, University of Arizona, Tucson, AZ 85721, USA}
\altaffiltext{8}{Steward Observatory, The University of Arizona, 933 N. Cherry Ave., Tucson, AZ 85721}
\altaffiltext{9}{Department of Physics \& Astronomy, Johns Hopkins University, 3400 N. Charles Street, Baltimore, MD 21218, USA}

\begin{abstract}

We present measurements of the luminosity and color-dependence of galaxy clustering at $0.2<z<1.0$ in the PRIsm MUlti-object Survey (PRIMUS).  
We quantify the clustering with the redshift-space and projected two-point correlation functions, $\xi(r_p,\pi)$ and $w_p(r_p)$, using volume-limited samples constructed from a parent sample of over $\sim130,000$ galaxies with robust redshifts in seven independent fields covering 9 ${\rm deg}^2$ of sky. 
We quantify how the scale-dependent clustering amplitude increases with increasing luminosity and redder color, with relatively small errors over large volumes. 
We find that red galaxies have stronger small-scale ($0.1<r_p<1~{\rm Mpc}/h$) clustering and steeper correlation functions compared to blue galaxies, as well as a strong color dependent clustering within the red sequence alone.
We interpret our measured clustering trends in terms of galaxy bias and obtain values of $b_{\rm gal}\approx0.9$-$2.5$, quantifying how galaxies are biased tracers of dark matter depending on their luminosity and color. 
We also interpret the color dependence with mock catalogs, and find that the clustering of blue galaxies is nearly constant with color, while redder galaxies have stronger clustering in the one-halo term due to a higher satellite galaxy fraction. 
In addition, we measure the evolution of the clustering strength and bias, and we do not detect statistically significant departures from passive evolution. 
We argue that the luminosity- and color-environment (or halo mass) relations of galaxies 
have not significantly evolved since $z\sim1$.  
Finally, using jackknife subsampling methods, we find that sampling fluctuations are important and that the COSMOS field is generally an outlier, due to having more overdense structures than other fields; we find that `cosmic variance' can be a significant source of uncertainty for high-redshift clustering measurements.  

\end{abstract}


\section{Introduction}



In the current paradigm of hierarchical structure formation, gravitational evolution causes dark matter particles to cluster around peaks of the initial density field and to collapse into virialized objects. 
These dark matter halos then provide the potential wells in which gas cools and galaxies subsequently form. 
In addition, there is a correlation between halo formation and abundances and the surrounding large-scale structure (Mo \& White 1996; Sheth \& Tormen 2002), while galaxy formation models assume that galaxy properties are determined by the properties of the host dark matter halo (Baugh et al.\ 1999; Benson et al.\ 2001).  
Therefore, correlations between halo properties and the environment induce observable correlations between galaxy properties and the environment. 

Correlations with large-scale structure are measured and quantified with a variety of techniques, including two-point correlation functions, which are the focus of this paper.   
Correlation function studies have shown that a variety of galaxy properties (such as luminosity, color, stellar mass, star formation rate, morphology, and spectral type) are environmentally dependent.  
In particular, luminous, red, massive, passively star-forming, and early-type galaxies have been found to be more strongly clustered than their (fainter, bluer, etc.) counterparts, and are hence more likely to reside in dense environments (e.g., Guzzo et al.\ 2000; Norberg et al.\ 2002; Madgwick et al.\ 2003; Zehavi et al.\ 2005; Skibba et al.\ 2009; de la Torre et al.\ 2011), and these correlations have been in place since at least $z\sim1$ (e.g, Coil et al.\ 2008; Quadri et al.\ 2008; Meneux et al.\ 2009), though they quantitatively exhibit substantial evolution with redshift. 

Such galaxy clustering analyses have been performed with galaxy redshift surveys at low redshift, such as the Sloan Digital Sky Survey (SDSS; York et al.\ 2000) and 2-degree Field Galaxy Redshift Survey (2dF; Colless et al.\ 2001), and at high redshift, such as the DEEP2 Galaxy Redshift Survey (Davis et al.\ 2003) and VIMOS-VLT Deep Survey (VVDS; Le F\'{e}vre et al.\ 2005).  The PRIsm MUlti-object Survey (PRIMUS; Coil et al.\ 2011; Cool et al.\ 2013) provides a `bridge' between these surveys, with hundreds of thousands of spectroscopic redshifts at $0.2<z<1$, allowing for the construction of volume-limited catalogs of faint galaxies with large dynamic range. 
PRIMUS is the first survey at $z>0.2$ to approach the volume and size of local surveys, and with greater depth. 
It is well-suited for clustering and other large-scale structure analyses, yielding new constraints on galaxy growth and evolution, and their connection to the assembly of dark matter (DM) halos.

Galaxy clustering is clearly correlated with luminosity and color, in a variety of wavelengths, in the nearby universe (Norberg et al.\ 2002; Zehavi et al.\ 2005, 2011; Tinker et al.\ 2008a; Skibba \& Sheth 2009), and luminosity and color dependent clustering has been studied at higher redshift as well (e.g., Pollo et al.\ 2006; Coil et al.\ 2008; Meneux et al.\ 2009; Abbas et al.\ 2010). 
Complementary to this work, clustering analyses at intermediate redshifts, between $z\sim0$ and $z\sim1$, are important for constraining analytic halo models and semi-analytic galaxy formation models.  
Some recent analyses are focused mainly on massive galaxies (e.g., Wake et al.\ 2008; H.\ Guo et al.\ 2013), while others use photometric redshifts and angular correlation functions (Brown et al.\ 2008; Ross et al.\ 2010; Coupon et al.\ 2012; Christodoulou et al.\ 2012), which are more difficult to interpret because of redshift uncertainties.  
Large samples of spectroscopic redshifts are necessary, and studies of fainter galaxies are needed as well, to complement those of more massive galaxies. 
The PRIMUS survey (and VIPERS\footnote{VIMOS Public Extragalactic Redshift Survey}; Guzzo et al.\ 2013) fulfills these requirements. 

For modeling and interpreting galaxy clustering trends, the 
halo model (see Cooray \& Sheth 2002; Mo, van den Bosch, \& White 2010 for reviews) has 
proven to be a useful framework. 
For example, such models have been used to interpret the luminosity and color dependence of galaxy clustering statistics (e.g., Zehavi et al.\ 2005; Phleps et al.\ 2006; Tinker et al.\ 2008a; Skibba \& Sheth 2009; Simon et al.\ 2009; Masaki et al.\ 2013; Hearin \& Watson 2013). 
The halo-model description of galaxy clustering is often done with the `halo occupation distribution' (HOD; Jing, Mo \& B\"orner 1998; Seljak 2000; Scoccimarro et al.\ 2001; Berlind \& Weinberg 2002; Kravtsov et al.\ 2004), 
which includes a prescription for the spatial distribution of `central' and `satellite' galaxies in halos as a function of halo mass. 
Recent analyses have built on this work with constraints on the evolution of halo occupation and the luminosity-halo mass relation (e.g., Conroy et al.\ 2006; Zheng et al.\ 2007; Leauthaud et al.\ 2012; Li et al.\ 2012; Yang et al.\ 2012). 
In this paper, we include some halo-model interpretations of luminosity and color dependent clustering in PRIMUS, while more sophisticated modeling of clustering as a function of stellar mass and star formation rate will be the focus of subsequent work. 


This paper is organized as follows. 
In the next section, we describe the PRIMUS survey, and the volume-limited catalogs we construct for the galaxy clustering measurements.
The galaxy clustering statistics and error analysis are described in Section~\ref{sec:methods}. 
In Sections~\ref{sec:LdepCFs} and \ref{sec:CdepCFs}, we present our luminosity and color dependent clustering results, including redshift-space and projected correlation functions. 
We present a halo-model interpretation of the results in Section~\ref{sec:model}, with galaxy bias and mock galaxy catalogs. 
Finally, we end with a discussion of our results in Section~\ref{sec:discussion}, 
including a discussion of results in the literature, galaxy evolution and clustering evolution, and cosmic variance. 

Throughout the paper we assume a spatially flat cosmology with 
$\Omega_m=0.27$ and $\Omega_\Lambda=0.73$, and $\sigma_8=0.8$, unless stated otherwise. 
These values of $\Omega_m$ and $\sigma_8$ are slightly lower than the latest cosmological constraints (Planck collaboration et al.\ 2013).
We write the Hubble constant as $H_0=100h$~km~s$^{-1}$~Mpc$^{-1}$. 
All magnitudes are based on the AB magnitude system (Oke \& Gunn 1983). 

\section{Data}

	\subsection{PRIMUS Galaxy Redshift Survey}\label{sec:primus}


The PRIMUS survey (Coil et al.\ 2011; Cool et al.\ 2013) 
is a spectroscopic faint galaxy redshift survey to $z \sim 1$ over seven fields on the sky. 
The survey covers 9.1 degrees to a depth of $i_{AB} \sim 23$.  
All objects in PRIMUS were observed with the IMACS spectrograph (Bigelow \& Dressler 2003) 
on the Magellan I Baade 6.5m telescope at Las Campanas Observatory in Chile. 
A low-dispersion prism and slitmasks are used to observe $\sim 2500$ objects at once 
with a field of view of 0.18 deg$^2$, and at each pointing, generally two slitmasks are used.  
PRIMUS targeted galaxies in a total of seven independent science fields: the 
Chandra Deep Field South-SWIRE (CDFS-SWIRE), the $02^{hr}$ and $23^{hr}$ DEEP2 fields, 
the COSMOS field, the European Large Area ISO Survey - South 1 field (ELAIS-S1; 
Oliver et al.\ 2000), the XMM-Large Scale Structure Survey field (XMM-LSS; Pierre et al.\ 2004), 
and the Deep Lens Survey (DLS; Wittman et al.\ 2002) F5 field.

Galaxy redshifts are obtained by fitting each spectrum with a galaxy template, 
and optical, \textit{GALEX}, and \textit{Spitzer} photometry are used to supplement the spectra as well as derive K-corrections (Moustakas et al.\ 2013; Cool et al.\ 2013). 
In total PRIMUS has $\sim$130,000 robust redshifts and a 
precision of $\sigma_z/(1+z)=0.005$ and to date, it is the largest intermediate-redshift faint galaxy survey. 

\subsection{Targeting Weights}\label{sec:weights}

Details of the PRIMUS target selection are given in Coil et al.\ (2011).  
Here we discuss the most salient points relevant for clustering measurements.  
The `primary' galaxy sample is defined as those galaxies that have a 
well-understood spatial selection function from which we can create a 
statistically complete sample.  As the footprint of our spectra on the 
detectors corresponds to an area of 30" by 8"  on the sky, any close pairs of 
galaxies can have only one galaxy of the pair observed on a given slitmask.  
While we observed two slitmasks per pointing to alleviate this problem, 
galaxies are sufficiently clustered in the plane of the sky such that even 
with two slitmasks we undersample the densest regions.  We therefore used a 
density-dependent selection weight, which tracked how many other galaxies 
would have had spectra that collided with the target galaxy, and selected a 
subsample of galaxies that would not overlap.  We thus avoided slit 
collisions and kept track of the known density-dependent targeting weight.  
By applying this known weight of each galaxy when calculating the 
correlation function, we can correct for this incompleteness in the data.

	\subsection{Volume-Limited Galaxy Catalogs}\label{sec:lumcats}

Throughout this paper, we select galaxies with the highest spectral quality ($Q=4$), which have the most confident redshifts.  
These redshifts have a typical precision of $\sigma_z/(1+z)=0.005$ and a $3\%$ outlier rate with respect to DEEP2, zCOSMOS (Lilly et al.\ 2007), and VVDS, with outliers defined as objects with $|\Delta z|/(1+z)>0.03$ (Coil et al.\ 2011; Cool et al.\ 2013).\footnote{We have tested with $Q\ge3$ galaxies (which have an outlier rate of $8\%$, and which would enlarge the samples by up to 30\% at high redshift) in Section~\ref{sec:LdepCFs} as well, and obtained approximately similar results, with measured correlation functions in agreement within $\approx15\%$ but with larger errors due to the larger redshift errors.  Robust redshifts are necessary for redshift- and magnitude-dependent sample selection and for measuring line-of-sight separations; for more discussion of effects of redshift errors and the robustness of clustering measurements, we refer the reader to Norberg et al.\ (2009), Zehavi et al.\ (2011), and Ross et al.\ (2012).}  
We use galaxies that are in the PRIMUS primary sample, which comprises objects with a recoverable spatial selection (see Coil et al.\ 2011 for details). 
Although PRIMUS covers a redshift range of $0.0\lesssim z \lesssim1.2$, for this study we use galaxies with redshifts $0.2<z<1.0$, to ensure complete samples with luminosities that can be compared over a wide redshift range. 

\begin{figure*}
	\includegraphics[width=1.0\linewidth]{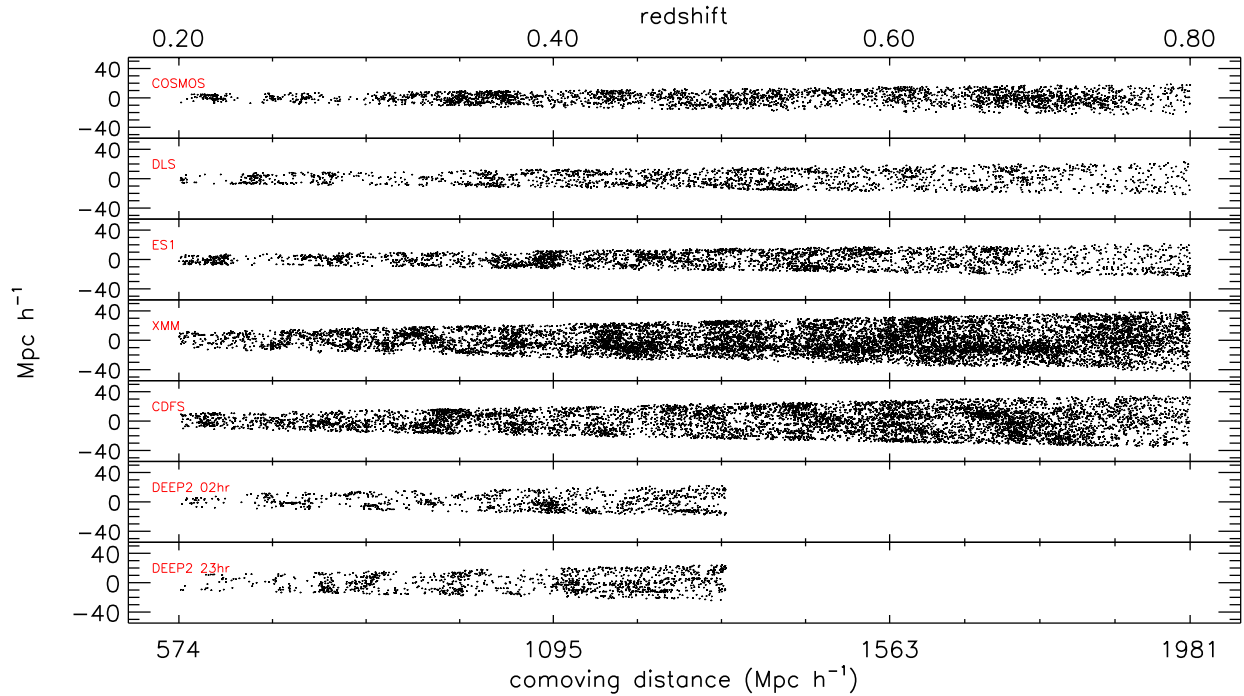} 
 	\caption{
        Redshift-space distribution of galaxies as a function of comoving distance along the line-of-sight and right ascension, relative to the median RA of the field. 
        From upper to lower panels, the corresponding fields are the following: COSMOS, DLS F5, ELAIS S1, XMM-LSS, CDFS-SWIRE, DEEP2 02hr, DEEP2 23hr. 
        Galaxies with $M_g<-17$ and high-quality redshifts ($Q=4$) are shown.}
 	\label{fig:cones}
\end{figure*}

Figure~\ref{fig:cones} shows the spatial distribution of galaxies in the seven science fields in comoving space.  Many large-scale structures and voids (underdense regions) are clearly visible, and the structures appear similar if a luminosity threshold (e.g., $M_g<-19$) is used.  
Note that we will exclude the $02^{hr}$ and $23^{hr}$ DEEP2 fields in the $0.5\lesssim z \lesssim1.0$ clustering analyses, 
as in these fields PRIMUS did not target galaxies above $z=0.65$, which already had spectroscopic redshifts in the DEEP2 galaxy redshift survey. 

Though large-scale structure can be seen in all of the fields in Figure~\ref{fig:cones}, in COSMOS (upper panel), there are several very overdense regions at $z\approx0.35$ and $z\approx0.7$, which have an impact on some of the clustering measurements. 
These have been previously identified by galaxy clustering and other methods (McCracken et al.\ 2007; Meneux et al.\ 2009; Kova\v{c} et al.\ 2010). 
The overdensity at $z\sim0.7$ appears to be due to a rich and massive cluster (Guzzo et al.\ 2007); such structures are rare and result in sampling fluctuations (e.g., Mo et al.\ 1992; Norberg et al.\ 2011), which we will discuss later in the error analyses and in Section~\ref{sec:var}.

\begin{figure}
   	\includegraphics[width=1.0\linewidth]{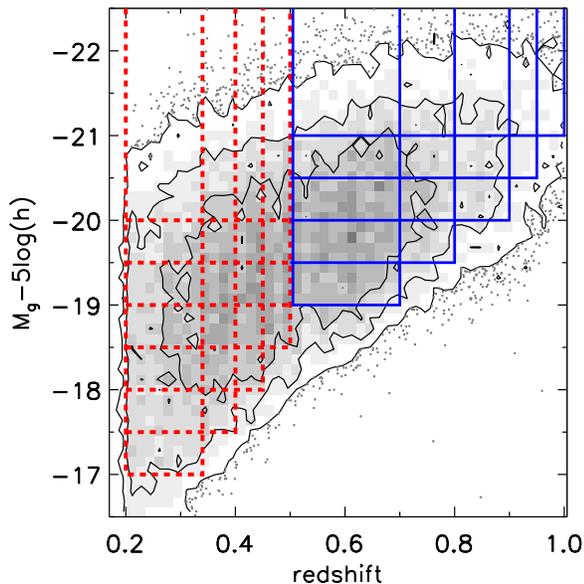} 
 		\caption{
                Contours of the galaxies used for this paper in $g$-band absolute magnitude and redshift space. 
                We divide the data into redshift bins at 0.2$<z<$0.5 (red lines) and 0.5$<z<$1.0 (blue lines) and construct volume-limited catalogs within those bins.  
                Details of the luminosity threshold and binned catalogs are given in Tables~\ref{tab:samples} and \ref{tab:samples2}.}
 	\label{fig:thresh_samples}
\end{figure}

From the flux-limited PRIMUS data set we create volume-limited samples in $M_g$ versus redshift-space (see Figure~\ref{fig:thresh_samples}). 
We divide the sample into the two redshift bins $0.2\lesssim z \lesssim0.5$ and $0.5\lesssim z \lesssim1.0$, which span roughly similar time-scales. 
We construct both luminosity-binned and threshold samples, which will be used for analyzing luminosity-dependent clustering in Section~\ref{sec:LdepCFs}.  Binned samples are one $M_g$ magnitude wide, while threshold samples are constructed for every half-magnitude step.

The luminosity-threshold volume-limited catalogs, including their galaxy numbers and number densities\footnote{Note that when calculating the number densities, we use in addition to the density-dependent weight the magnitude weight (see Coil et al.\ 2011 and Moustakas et al.\ 2013 for details).}, are described in Tables~\ref{tab:samples} and \ref{tab:samples2}.  
%
For reference, the $g$-band $M_\ast$, corresponding to the $L_\ast$ characteristic luminosity of luminosity functions (LFs, which are fitted to a Schechter function), is approximately $M_\ast\approx-20.4\pm0.1$ at $z\sim0.5$, based on the SDSS, GAMA\footnote{Galaxy and Mass Assembly (Driver et al.\ 2011).}, AGES\footnote{AGN and Galaxy Evolution Survey (Kochanek et al.\ 2012).}, and DEEP2 LFs (Blanton et al.\ 2003; Loveday et al.\ 2012; Cool et al.\ 2012; Willmer et al.\ 2006). 

\begin{table}
\caption{Limits and Number Densities of Volume-limited Catalogs with Luminosity Thresholds}
 \centering
 \begin{tabular}{ l | c c c c c c }
   \hline
    $M_g^\mathrm{max}$ & $\langle M_g\rangle$ & $\langle z\rangle$ & $z_{\rm min}$ & $z_{\rm max}$ & $N_{\rm gal}$ & $\bar n_{\rm gal, wt}$ \\
   \hline
   -17.0 & -18.79 & 0.28 & 0.20 & 0.34 & 9576 & 5.59 \\
   -17.5 & -19.05 & 0.32 & 0.20 & 0.40 & 14078 & 4.81 \\ 
   -18.0 & -19.28 & 0.35 & 0.20 & 0.45 & 16671 & 3.98 \\
   -18.5 & -19.54 & 0.38 & 0.20 & 0.50 & 17465 & 3.07 \\
   -19.0 & -19.81 & 0.38 & 0.20 & 0.50 & 13158 & 2.34 \\
   -19.5 & -20.13 & 0.39 & 0.20 & 0.50 & 8665 & 1.55 \\
   -20.0 & -20.47 & 0.39 & 0.20 & 0.50 & 4817 & 0.88 \\
   \hline 
   -19.0 & -20.10 & 0.60 & 0.50 & 0.70 & 11717 & 1.73 \\ 
   -19.5 & -20.37 & 0.64 & 0.50 & 0.80 & 12998 & 1.17 \\
   -20.0 & -20.68 & 0.68 & 0.50 & 0.90 & 11365 & 0.71 \\ 
   -20.5 & -21.02 & 0.71 & 0.50 & 0.95 & 7285 & 0.40 \\ 
   -21.0 & -21.40 & 0.74 & 0.50 & 1.00 & 3818 & 0.19 \\ 
   \hline
  \end{tabular}
  \begin{list}{}{}
    \setlength{\itemsep}{0pt}
    \item Luminosity threshold catalogs: limits, mean luminosity and redshift, number counts, and weighted number densities (in units of $10^{-2}h^3\mathrm{Mpc}^{-3}$, using weights described in Sec.~\ref{sec:weights}) for galaxies with $Q=4$ redshifts in the PRIMUS fields.  (See text for details.) 
  \end{list}
\label{tab:samples}
\caption{Limits and Number Densities of Volume-limited Catalogs with Luminosity Bins}
 \centering
 \begin{tabular}{ c c | c c c c c c }
   \hline
    $M_g^\mathrm{max}$ & $M_g^\mathrm{min}$ & $\langle M_g\rangle$ & $\langle z\rangle$ & $z_{\rm min}$ & $z_{\rm max}$ & $N_{\rm gal}$ & $\bar n_{\rm gal, wt}$ \\
   \hline
   -17.0 & -18.0 & -17.55 & 0.27 & 0.20 & 0.34 & 2325 & 1.39 \\ 
   -18.0 & -19.0 & -18.53 & 0.34 & 0.20 & 0.45 & 6560 & 1.54 \\ 
   -19.0 & -20.0 & -19.47 & 0.38 & 0.20 & 0.50 & 8341 & 1.46 \\ 
   -20.0 & -21.0 & -20.39 & 0.39 & 0.20 & 0.50 & 4220 & 0.77 \\ 
   \hline 
   -19.0 & -20.0 & -19.54 & 0.60 & 0.50 & 0.70 & 5477 & 0.79 \\ 
   -20.0 & -21.0 & -20.45 & 0.67 & 0.50 & 0.90 & 8179 & 0.50 \\ 
   -21.0 & -22.0 & -21.34 & 0.74 & 0.50 & 1.00 & 3488 & 0.17 \\ 
   \hline
  \end{tabular}
  \begin{list}{}{}
    \setlength{\itemsep}{0pt}
    \item Luminosity-binned catalogs: limits, numbers, and number densities (in units of $10^{-2}h^3\mathrm{Mpc}^{-3}$) for galaxies with $Q=4$ redshifts in the PRIMUS fields. 
  \end{list}
\label{tab:samples2}
\end{table}

\subsection{Color-Dependent Galaxy Catalogs}\label{sec:colcats}

We now describe how the color-dependent catalogs are constructed, which are used for analyzing color dependence of galaxy clustering in Section~\ref{sec:CdepCFs}.  

\begin{figure}
   	\includegraphics[width=1.0\linewidth]{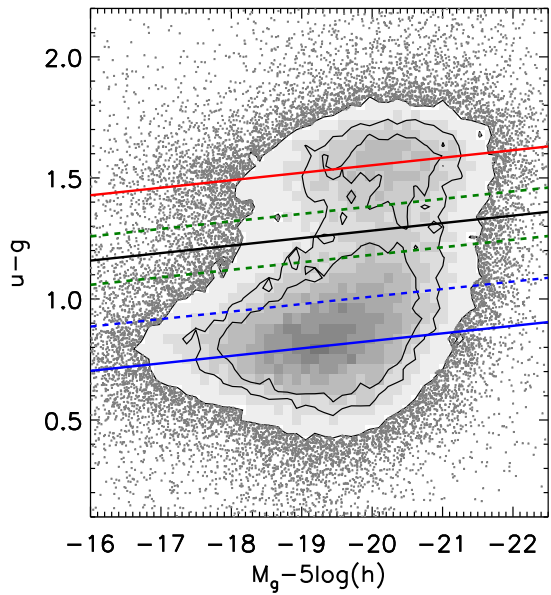} 
 	\caption{ Contours of galaxies in the $u-g$ color-magnitude diagram, where 
primary galaxies with high-quality redshifts in the range $0.2<z<1.0$ are shown. 
        The black line indicates the division between red and blue galaxies, using Eqn.~(\ref{eq:colcut}), and the red, blue, and green lines demarcate the finer color bins (Eqns.~\ref{eqn:finecolcuts}; $z=0.5$ is used as the reference redshift here). 
        }
	\label{fig:color_samples}
\end{figure}

We begin with the PRIMUS $(u-g)-M_g$ color-magnitude distribution (CMD), defined with 
$u$- and $g$-band magnitudes, which is shown in Figure~\ref{fig:color_samples}. 
The distribution of $p(u-g|M_g)$ is clearly bimodal, and can be approximately described with a double-Gaussian distribution at fixed luminosity (e.g., Baldry et al.\ 2004; Skibba \& Sheth 2009).  
We simply separate the `blue cloud' and `red sequence' using the minimum between these modes, which approximately corresponds to the following red/blue division (see also Aird et al.\ 2012): 
\begin{equation}
 (u-g)_{\rm cut}  = -0.031M_g - 0.065z + 0.695
 \label{eq:colcut}
\end{equation}
We define a `green valley' (e.g., Wyder et al.\ 2007; Coil et al.\ 2008) component as well, within 0.1~mag of this red/blue demarcation.  
Such galaxies are often interpreted as in transition between blue and red galaxies (but see Schawinski et al.\ 2013), and we can determine whether their clustering strength lies between that of their blue and red counterparts (see Sec.~\ref{sec:redblue}).

\begin{table*}
\caption{Limits and Number Densities of Volume-limited Catalogs of Blue and Red Galaxies}
 \centering
 \begin{tabular}{ c c | c c c c | c c c c }
   \hline
    $M_g^\mathrm{max}$ & $\langle z\rangle$ & $\langle M_g\rangle_{\rm blue}$ & $\langle u-g\rangle_{\rm blue}$ & blue $N_{\rm gal}$ & blue $\bar n_{\rm gal, wt}$ & $\langle M_g\rangle_{\rm red}$ & $\langle u-g\rangle_{\rm red}$ & red $N_{\rm gal}$ & red $\bar n_{\rm gal, wt}$\\
   \hline
   -17.0 & 0.28 & -18.54 & 0.87 & 6667  & 3.78 & -19.31 & 1.53 & 2909 & 1.80 \\
   -17.5 & 0.32 & -18.80 & 0.89 & 9447  & 3.11 & -19.44 & 1.54 & 4631 & 1.69 \\
   -18.0 & 0.35 & -19.05 & 0.90 & 10892 & 2.51 & -19.60 & 1.54 & 5779 & 1.47 \\
   -18.5 & 0.38 & -19.35 & 0.92 & 10918 & 1.85 & -19.76 & 1.55 & 6547 & 1.22 \\
   -19.0 & 0.38 & -19.65 & 0.95 & 7599  & 1.30 & -19.92 & 1.55 & 5559 & 1.04 \\
   -19.5 & 0.39 & -20.01 & 0.99 & 4619  & 0.79 & -20.18 & 1.56 & 4046 & 0.76 \\
   -20.0 & 0.39 & -20.38 & 1.02 & 2323  & 0.40 & -20.49 & 1.57 & 2494 & 0.48 \\
   \hline 
   -19.0 & 0.60 & -19.89 & 0.89 & 7167  & 1.06 & -20.33 & 1.59 & 4550 & 0.67 \\
   -19.5 & 0.64 & -20.19 & 0.90 & 7407  & 0.67 & -20.52 & 1.59 & 5591 & 0.50 \\
   -20.0 & 0.68 & -20.52 & 0.93 & 5915  & 0.37 & -20.76 & 1.59 & 5450 & 0.34 \\
   -20.5 & 0.71 & -20.92 & 0.97 & 3356  & 0.18 & -21.04 & 1.60 & 3929 & 0.22 \\
   -21.0 & 0.74 & -21.32 & 1.00 & 1553  & 0.08 & -21.37 & 1.62 & 2265 & 0.11 \\
   \hline
  \end{tabular}
  \begin{list}{}{}
    \setlength{\itemsep}{0pt}
    \item Luminosity threshold catalogs of red and blue galaxies: limits, numbers, and number densities (in units of $10^{-2}h^3\mathrm{Mpc}^{-3}$) for galaxies with $Q=4$ redshifts in the PRIMUS fields.  See Sec.~\ref{sec:colcats} for details and Sec.~\ref{sec:CdepCFs} for results.
  \end{list}
\label{tab:samples3}
\end{table*}
\begin{table}
\caption{Limits and Number Densities of Volume-limited Catalogs with Color Bins}
 \centering
 \begin{tabular}{ l | c c c c c }
   \hline
    name & $\langle M_g\rangle$ & $\langle z\rangle$ & $\langle u-g\rangle$ & $N_{\rm gal}$ & $\bar n_{\rm gal, wt}$ \\
   \hline
   bluest    & -19.78 & 0.53 & 0.695 & 6150 & 0.32 \\
   bluecloud & -19.89 & 0.52 & 0.911 & 6229 & 0.31 \\
   bluer     & -20.13 & 0.52 & 1.135 & 6103 & 0.31 \\
   green     & -20.20 & 0.52 & 1.288 & 3743 & 0.20 \\
   redder    & -20.22 & 0.53 & 1.440 & 6314 & 0.34 \\
   reddest   & -20.27 & 0.57 & 1.685 & 6341 & 0.36 \\
   \hline
  \end{tabular}
  \begin{list}{}{}
    \setlength{\itemsep}{0pt}
    \item Color-binned catalogs: properties, numbers, and number densities (in units of $10^{-2}h^3\mathrm{Mpc}^{-3}$) for galaxies with $Q=4$ redshifts in the PRIMUS fields.  
    All of the color-binned catalogs have $M_g<-19.0$ and $0.20<z<0.80$. 
  \end{list}
\label{tab:samples4}
\end{table}

In addition, to analyze the clustering dependence as a function of color, we use finer color bins.  
For these, we slice the color-magnitude distribution with lines parallel to the red/blue demarcation. 
(It is perhaps more accurate to use a steeper blue sequence cut, as the blue sequence has a steeper luminosity dependence than the red one, but we find that this choice does not significantly affect our results.) 
The blue cloud is divided into three catalogs, while the red sequence is divided into two, 
and the cuts are chosen to select color-dependent catalogs with an approximately similar number for a luminosity threshold of $M_g\geq-19$.
In particular, we apply the following color cuts:
\begin{eqnarray}
   (u-g)_{\rm red}   &=& -0.031 M_g - 0.065z + 0.965 \nonumber\\
   (u-g)_{\rm blue1} &=& -0.031 M_g - 0.12z + 0.45 \label{eqn:finecolcuts}\\
   (u-g)_{\rm blue2} &=& -0.031 M_g - 0.12z + 0.267 \nonumber
\end{eqnarray}

A potentially important caveat is that 
photometric offsets between the fields (as the restframe colors in each field are interpolated from the observed photometry using \texttt{kcorrect}; Blanton \& Roweis 2007) and uncertainties in the targeting weights result in the CMDs not being entirely identical across the PRIMUS fields. 
In order to address this, we assign different color-magnitude cuts to each field, based on their $p(c|L)$ distributions (i.e., their color distributions as a function of luminosity), while ensuring that each of the color fractions are similar. 
The redshift dependence of the cuts is based on the approximate redshift evolution of the red sequence and blue cloud (Aird et al.\ 2012), and is not varied among the fields. 
For our results with the finer color bins (Section~\ref{sec:finecol}), we take this approach and proportionally split each field separately (using cuts very similar to Eqns.~\ref{eqn:finecolcuts}), but we find that strictly applying the same cuts to each field yields nearly the same results (the resulting color-dependent correlation functions differ by at most $10\%$).  

The catalogs of red and blue galaxies and finer color bins are described in Tables~\ref{tab:samples3} and \ref{tab:samples4}.

\section{Galaxy Clustering Methods}\label{sec:methods}

	\subsection{Two-Point Correlation Function}\label{sec:2PCF}


The two-point autocorrelation function $\xi(r)$ is a powerful tool to characterize galaxy clustering, by quantifying the excess probability $dP$ over random of finding pairs of objects as a function of separation (e.g., Peebles 1980). 
That is,
\begin{equation}
dP\,=\,n[1+\xi(r)]dV ,
\end{equation}
where $n$ is the number density of galaxies in the catalog.  

To separate effects of redshift distortions and spatial correlations, we estimate the correlation function on a two-dimensional grid of pair separations parallel ($\pi$) and perpendicular ($r_p$) to the line-of-sight.  
Following Fisher et al.\ (1994), we define vectors ${\bf v}_1$ and ${\bf v}_2$ to be the redshift-space positions of a pair of galaxies, ${\bf s}$ to be the redshift-space separation $({\bf v}_1-{\bf v}_2)$, and ${\bf l}=({\bf v}_1+{\bf v}_2)/2$ to be the mean coordinate of the pair.  The parallel and perpendicular separations are then 
\begin{equation}
\pi\equiv|{\bf s} \cdot {\bf l}|/|{\bf l}|, \qquad r_p^2\equiv {\bf s} \cdot {\bf s} - \pi^2 .
\end{equation} 

We use the Landy \& Szalay (1993) estimator
\begin{equation}
\xi_i(r_p,\pi)=\frac{DD(r_p,\pi)-2DR(r_p,\pi)+RR(r_p,\pi)}{RR(r_p,\pi)} ,
\label{eq:LS}
\end{equation}
where $DD$, $DR$, and $RR$ are the counts of data-data, data-random, and random-random galaxy pairs, respectively,  as a function of $r_p$ and $\pi$ separation, in the field $i$.  
The $DD$ and $DR$ pair counts are accordingly weighted by the total targeting weights (named `targ\_weight'; see Sec.~\ref{sec:weights}). 
$DD$, $DR$, and $RR$ are normalized by $n_D(n_D-1)$, $n_D n_R$, and $n_R(n_R-1)$, respectively, where $n_D$ and $n_R$ are the mean number densities of the data and random catalogs (the randoms are described in Sec.~\ref{sec:randoms}).  
We have tested and verified that this estimator (\ref{eq:LS}) yields clustering results that are nearly identical to those with other estimators (including Hamilton 1993 and $DD/RR-1$; see also Kerscher et al.\ 2000 and Zehavi et al.\ 2011). 

Because we have multiple fields that contribute to a composite PRIMUS correlation function, we compute a correlation function for each field and weight by the number of galaxies in that field divided by the total number of objects in all the fields combined. 
This can be written as the following:
\begin{equation}
\xi (r_p,\pi) = \frac{\displaystyle\sum\limits_{i=0}^{n_{\rm field}} N_{d,i} \left(DD_i - 2DR_i + RR_i \right)}{\displaystyle\sum\limits_{i=0}^{n_{\rm field}} N_{d,i} RR_i} ,
\label{eq:xifield1}
\end{equation}
which is similar, but not equivalent, to:   
\begin{equation}
\xi (r_p,\pi) = \frac{1}{N_{d,tot}}\displaystyle\sum\limits_{i=0}^{n_{\rm field}} N_{d,i}\; \xi_i
\label{eq:xifield2}
\end{equation}
where $N_{d,i}$ is the number of galaxies in the $i^{\rm th}$ field. 
In this way, the larger fields (where the signal to noise is higher) contribute more than the smaller fields.  In practice, we evaluate the former expression (Eqn.~\ref{eq:xifield1}) for the composite correlation function, though Eqn.~(\ref{eq:xifield2}) is nearly identical. 

To recover the real-space correlation function $\xi(r)$, we integrate $\xi(r_p,\pi)$ over the $\pi$ direction since redshift-space distortions are only present along the line-of-sight direction. The result is the projected correlation function, which is defined as
\begin{equation}
w_p(r_p)=2\int_0^{\infty}d\pi\, \xi(r_p,\pi) 
    = 2\int_{r_p}^\infty dr\,\frac{r\,\xi(r)}{\sqrt{r^2-r_p^2}} 
\label{eq:wp}
\end{equation}
(Davis \& Peebles 1983). 
 If we assume that $\xi(r)$ can be represented by a power-law, $(r/r_0)^{-\gamma}$, then the analytic solution to Eqn.~(\ref{eq:wp}) is
\begin{equation}
w_p(r_p) = r_p\left( \frac{r_0}{r_p} \right) ^{\gamma} \frac{\Gamma(1/2)\Gamma[(\gamma-1)/2]}{\Gamma(\gamma/2)}
 \label{eq:powerlaw}
\end{equation}

\subsection{Construction of Random Catalogs}\label{sec:randoms}


For each PRIMUS field, a random catalog is constructed with a survey geometry and angular selection function similar to that of the data field and with a redshift distribution modeled by smoothing the data field redshift distribution. 
Each random catalog contains 25-40 times as many galaxies as its corresponding field (to limit Poisson errors in the measurements), depending on the varying number density and size of the sample.  
We have verified that increasing the number of random points has a negligible effect on the measurements, and other studies have found that random catalogs of this size are sufficient to minimize Poisson noise at the galaxy separations we consider (Zehavi et al.\ 2011; Vargas-Maga\~na et al.\ 2013).  

In addition to the targeting weights discussed above (Sec.~\ref{sec:weights}), redshift confidence weights are needed 
because redshift completeness varied slightly across the sky, due to observing conditions on a given slitmask. 
To account for this, we use the 
\texttt{mangle}\footnote{\texttt{http://space.mit.edu/{\raise.17ex\hbox{$\scriptstyle\sim$}}molly/mangle}} pixelization algorithm (Swanson et al.\ 2008b) to divide the individual fields into areas of $\sim0.01$ deg$^2$ on the plane of the sky.  
In these smaller regions, we then find the ratio of the number of $Q=4$ galaxies to all galaxies and use this number to upweight our random catalogs accordingly, 
though we excluded regions in which the redshift success rate was particularly low. 
We have also used simple mock catalogs to test the PRIMUS mask design and compare the measured correlation functions to those recovered using the observed galaxies and target weights, from which we find no systematic effects due to target sampling. 
We convert the coordinates of each galaxy from (RA, Dec, $z$) to the comoving coordinate ($r_x$,$r_y$,$r_z$) space using the \texttt{red} program\footnote{L.~Moustakas and J.~Moustakas, http://code.google.com/p/red-idl-cosmology/}. 


The total redshift distribution, $N(z)$, of galaxies in PRIMUS with robust redshifts is fairly smooth (see Coil et al.\ 2011).  
Nonetheless, $N(z)$ varies significantly among the PRIMUS fields and for different luminosity thresholds and bins, and can be much less smooth than the combined $N(z)$, due to large-scale structure.  
One approach is to randomly shuffle the redshifts in $N(z)$ (see discussion in Ross et al.\ 2012), or one can fit a smooth curve to the distribution for the different pointings. 
We choose the latter and smooth the luminosity-dependent redshift distributions for each field $i$, $N_i(z|L)$, and use this for the corresponding random catalogs.  
From tests of $N(z)$ models and smoothing methods, we find that this choice and the 
choice of smoothing parameters affect the correlation functions by a few per cent at small scales and up to $\sim20\%$ at large scales ($r\geq10~{\rm Mpc}/h$).

\subsection{Measuring Correlation Functions}\label{sec:CFmeas}


Most of the clustering analysis in this paper is focused on measuring and interpreting projected correlation functions, $w_p(r_p)$ (Eqn.~\ref{eq:wp}), which are obtained by integrating $\xi(r_p,\pi)$.  
In practice, we integrate these out to $\pi_{\rm max}=80~{\rm Mpc}/h$, which is a scale that includes most correlated pairs while not adding noise created by uncorrelated pairs at larger separations along the line-of-sight. Bins are linearly spaced in the $\pi$ direction with widths of $5~{\rm Mpc}/h$. 
The use of a finite $\pi_{\rm max}$ means that the correlation functions suffer from residual redshift-space distortions, but from the analysis of van den Bosch et al.\ (2013), we expect these to be on the order of $10\%$ at $r_p>10~{\rm Mpc}/h$.  
After performing many tests of $w_p(r_p)$ measurements over the PRIMUS redshift range, we find that values of $50<\pi_{\rm max}<100~{\rm Mpc}/h$ produce robust clustering measurements that are not significantly dependent on this parameter.

We will present correlation functions as a function of luminosity, color, and redshift in Sections~\ref{sec:LdepCFs} and \ref{sec:CdepCFs}. 
We will also fit power-laws to the correlation functions at large scales ($0.5<r_p<10~{\rm Mpc}/h$).  
However, there are small deviations from a power-law form, due to galaxy pairs in single dark matter halos and in separate halos. These are referred to as the `one-halo' and `two-halo' terms and overlap at $r_p\sim1$-$2~{\rm Mpc}/h$ (Zehavi et al.\ 2004; Watson et al.\ 2011). 
In addition, we will estimate the galaxy bias at large separations, which quantifies the galaxies' clustering strength with respect to DM (e.g., Berlind \& Weinberg 2002). 


\subsection{Error Estimation}\label{sec:errors}

For our error analyses, we use `internal' error estimates, methods using the dataset itself. 
This involves dividing our galaxy catalogs into subcatalogs. 
We do this by cutting the large fields (XMM and CDFS) along RA and Dec, and requiring that the subcatalogs have approximately equal area (within 20\%) and are sufficiently large for the clustering measurements (see also Bray et al., in prep.). 

Except when stated otherwise, we use `jackknife' errors (e.g., Lupton 1993; Scranton et al.\ 2002), which are estimated as follows: 
\begin{equation}
 [\Delta w(r_p)]^2 \,=\, \frac{N_{\rm jack}-1}{N_{\rm jack}} \sum_{i=1}^{N_{\rm jack}} [w_i(r_p)-{\bar w}_i(r_p)]^2 ,
\label{eq:jack}
\end{equation}
with $N_{\rm jack}=9$ or 11 resamplings, as the two DEEP2 fields are included only in the low-redshift bins. 
However, Eqn.~(\ref{eq:jack}) is designed for Gaussian statistics, and some of our measured clustering statistics have outliers (which are discussed below), so rather than using standard deviations, we use the 16 and 84 percentiles (and interpolate between measurements when necessary). 
In addition, we have tested these errors with mock catalogs (described in Sec.~\ref{sec:mock}), in which we obtained approximately consistent errors except at large scales ($r_p>5~{\rm Mpc}/h$), where the mock errors were smaller than the PRIMUS ones, primarily due to the mocks' larger volumes.

For comparison, we also compute (block) bootstrap errors, which involves resampling with replacement (Barrow et al.\ 1984; Loh 2008).  These errors are estimated with the following: 
\begin{equation}
 [\Delta w(r_p)]^2 \,=\, \frac{1}{N_{\rm boot}-1} \sum_{i=1}^{N_{\rm boot}} [w_i(r_p)-{\bar w}_i(r_p)]^2 ,
\label{eq:boot}
\end{equation}
with $N_{\rm boot}=100$ resamplings. 
For an analysis of the robustness of various jackknife and bootstrap clustering error estimates, we refer the reader to Norberg et al.\ (2009, 2011).

In Section~\ref{sec:LdepCFs}, we find that the jackknife errors of $w_p(r_p)$ (and quantities inferred from it) do not have a Gaussian distribution, often due to the different clustering in the COSMOS field, which produces outliers.  
We therefore use the 16 and 84 percentiles for the error bars, which we argue are the most robust for these measurements, rather than the standard deviations (Eqn.~\ref{eq:jack}), which yield overestimated errors in comparison: the 16 and 84 percentiles result in errors that are $\sim15-20\%$ smaller than these.  
Bootstrap errors (Eqn.~\ref{eq:boot}) allow for more resamplings, and these yield errors smaller by $\sim30\%$ or less. 
In addition, we have measured these correlation functions for $Q\ge3$ redshifts as well (described in Sec.~\ref{sec:lumcats}), which result in clustering measurements that are similar or at slightly lower amplitude (by $<15\%$) and larger errors (by $\sim20\%$) than the measurements with $Q=4$ errors.

We also attempt to account for the effects of uncertainty in the redshift distributions and residual redshift-space distortions (described in Sec.~\ref{sec:randoms} and \ref{sec:CFmeas}); these contributions are relatively small, though they contribute up to $20\%$ error at the largest separations. 
Lastly, note that for the power-law fits to the correlation functions (in Sec.~\ref{sec:LdepCFs} and \ref{sec:CdepCFs}), we use the clustering measurements of all of the subsamples, rather than the covariance matrices, which we find to be too noisy.  In particular, we calculate the power-law parameters ($r_0$ and $\gamma$) for each jackknife subsample and measure their variance to obtain the error.

\section{Results: Luminosity-dependent clustering}\label{sec:LdepCFs}

\subsection{Redshift-space Clustering: $\xi(r_p,\pi)$}\label{sec:xi}

We begin with correlation functions as a function of projected ($r_p$) and line-of-sight 
($\pi$) separation, which were described in Section~\ref{sec:2PCF}. 
These redshift-space correlation functions, $\xi(r_p,\pi)$, have been previously shown to depend on galaxy luminosity (e.g., Hawkins et al.\ 2003; Skibba et al.\ 2006); they vary with color as well (e.g., Zehavi et al.\ 2005; Coil et al.\ 2008; Loh et al.\ 2010), which we investigate in Section~\ref{sec:CdepCFs}. 

\begin{SCfigure}
        \includegraphics[width=0.54\hsize]{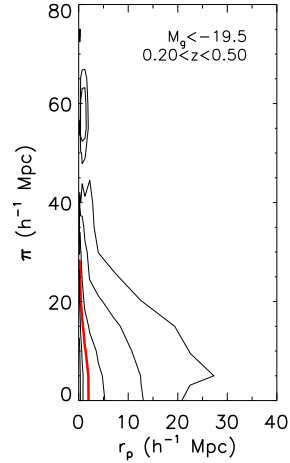}
 	\caption{
          Redshift-space two-dimensional correlation function $\xi(r_p,\pi)$ (smoothed here for clarity using a $5\times5$ h$^{-1}$Mpc boxcar technique) for all $Q=4$ galaxies in the volume-limited threshold sample with $M_{\mathrm{g}}<-19.5$ and $0.20<z<0.50$.  
	  The contour levels are 0.2, 0.5, 1.0 (thick red line), 2.0, and 5.0. 
          For the projected correlation functions (below), we integrate out to $\pi_{\rm max}=80~{\rm Mpc}/h$.
          Redshift-space distortions are clearly present on small scales ($r_p<$few~Mpc$/h$), and they are dominated by FOG, while redshift errors contribute only a small amount (see text for details). 
        } 
 	\label{fig:xirppi}
\end{SCfigure}

$\xi(r_p,\pi)$ in PRIMUS at low redshifts ($z<0.5$) for the $M_g<-19.5$ sample is shown in Figure~\ref{fig:xirppi}.  
This clearly demonstrates the effects of redshift-space distortions, with the so-called `fingers-of-god' (FOG; Jackson 1972; Peebles 1980), the elongated clustering along the line-of-sight at small separations, due to the virial motions of galaxies within halos; and at large separations, the compression in the $\pi$ direction due to coherent large-scale streaming (Kaiser 1987).\footnote{Note that these two effects, the small-scale FOG and large-scale squashing effect, are not independent (Scoccimarro 2004).} 


Although the small-scale redshift-space distortions are dominated by FOG, in the presence of redshift uncertainties the FOG will appear larger than in the absence of them.  
This is the case at $z>0.5$ (not shown), where the PRIMUS redshift uncertainties $\sigma_z/(1+z)$ are larger. 
We have tested this by adding Gaussian error to the redshifts in low-$z$ samples, and remeasured $\xi$, obtaining larger and smeared out FOG distortions.  Similarly, in a mock galaxy catalog (see Sec.~\ref{sec:mock} for details) without redshift errors, the FOGs are more distinct and slightly smaller. 
We estimate that the small-scale FOG distortions are extended by 5~Mpc/$h$ or more by the redshift errors at $z>0.5$, while the large-scale redshift-space clustering signal is only slightly reduced\footnote{Note that redshift errors significantly affect high-$z$ galaxy environment measures as well (Shattow et al.\ 2013).}; the effect of redshift errors are much smaller than this at $z<0.5$.  
Although the correlation function for the galaxies with high quality redshifts ($Q=4$, see Sec.~\ref{sec:lumcats}) is shown in the figure, the result for $Q\ge3$ is nearly identical, confirming the small effect of the redshift errors. 
In addition, our $\xi(r_p,\pi)$ measurements here demonstrate that the choice of $\pi_{\rm max}=80~{\rm Mpc}/h$ is sufficient to integrate over the FOG distortions (including in the higher redshift samples). 


\subsection{Projected Clustering: $w_p(r_p)$}\label{sec:wp}

\begin{figure*}
	\includegraphics[width=0.497\hsize]{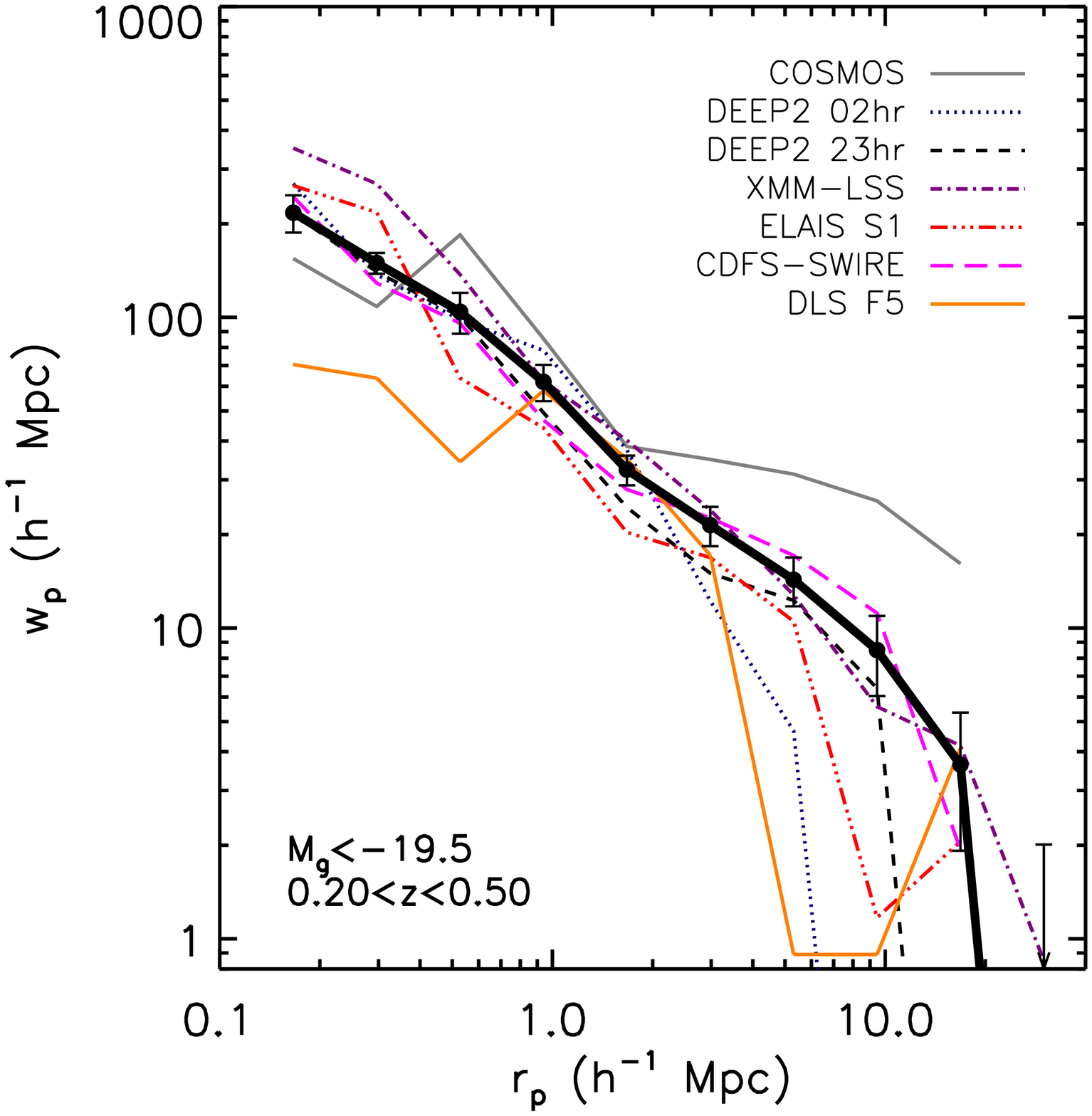} 
	\includegraphics[width=0.497\hsize]{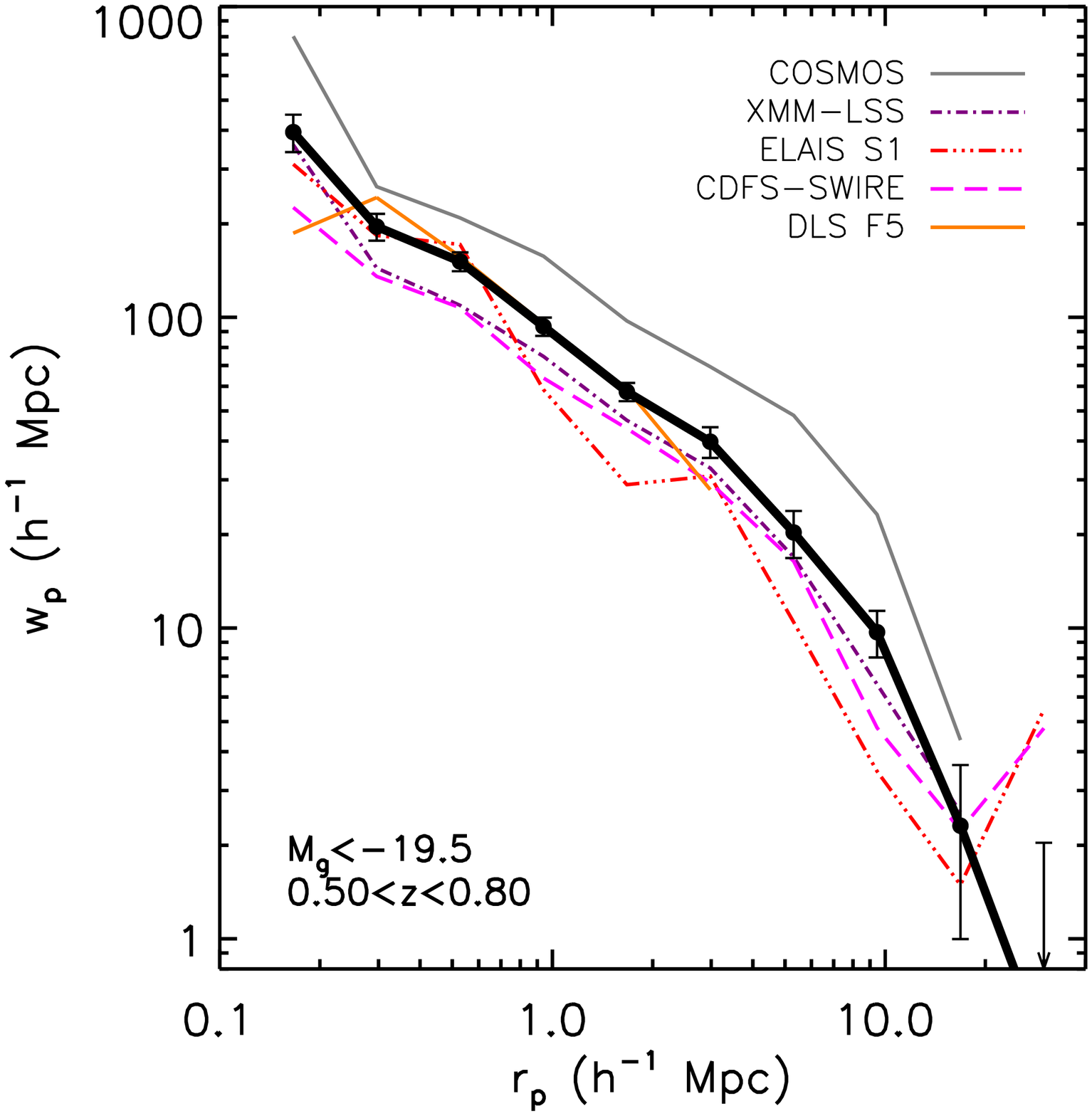} 
 	\caption{
         The projected correlation function of galaxies in each individual field in the volume-limited threshold samples for $M_g<-19.5$ at low redshift (left) and high redshift (right). 
         The thick, solid line is the coadded projection correlation function of the underlying fields, where each field is weighted according to the number of PRIMUS galaxies in the particular sample. 
         While there is substantial variation among the fields, the composite correlation function (thick black line) is smooth and well-behaved.  Its errors are estimated with jackknife subsampling (see Sec.~\ref{sec:errors}). 
        }

 \label{fig:wpindfields}
\end{figure*}

Next, we present the projected correlation functions, $w_p(r_p)$, of individual PRIMUS fields for $M_g<-19.5$ in Figure~\ref{fig:wpindfields}. 
There is clearly considerable variation in the clustering signal of the PRIMUS fields, at all separations, which is an expected effect of `cosmic variance' (or more accurately, sampling fluctuations), such that field-to-field variation in excess of shot noise are found in finite volume surveys, due to large-scale structures (e.g., Diaferio et al.\ 1999; Somerville et al.\ 2004). 
Note that the two DEEP2 fields are the smallest PRIMUS fields and cannot probe large galaxy separations. 
The composite correlation function of all the fields is dominated by the largest ones, XMM and CDFS, 
and these two fields have consistent and smoothly varying correlation functions.  
Note that the clustering in COSMOS is stronger than in the other fields, especially at high redshift, which is likely due to particularly large structures in this field, as noted above.  
In contrast, for example, the high-$z$ ELAIS-S1 field has a weaker clustering signal. 

\begin{figure*}
	\includegraphics[width=0.497\hsize]{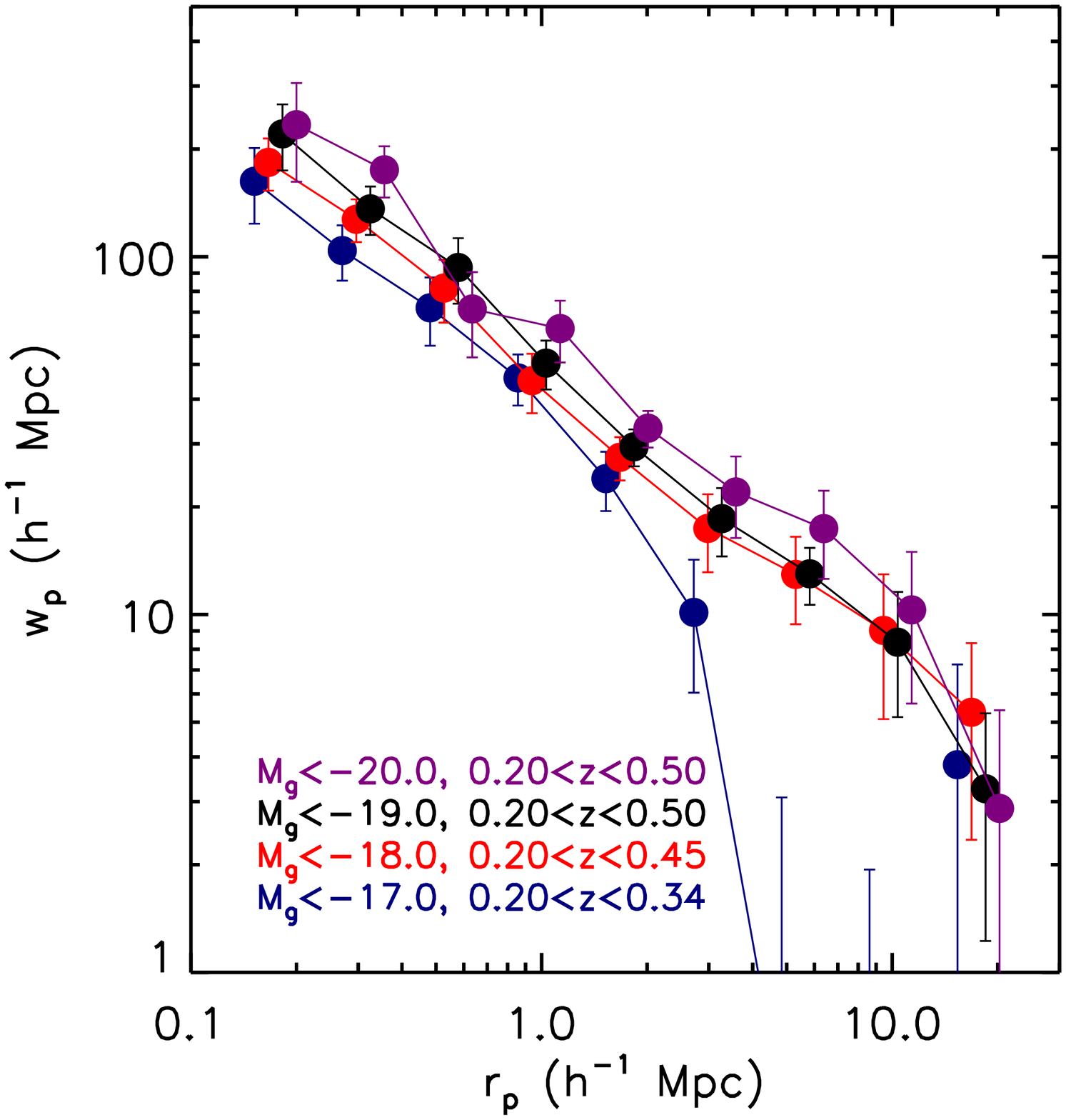} 
	\includegraphics[width=0.497\hsize]{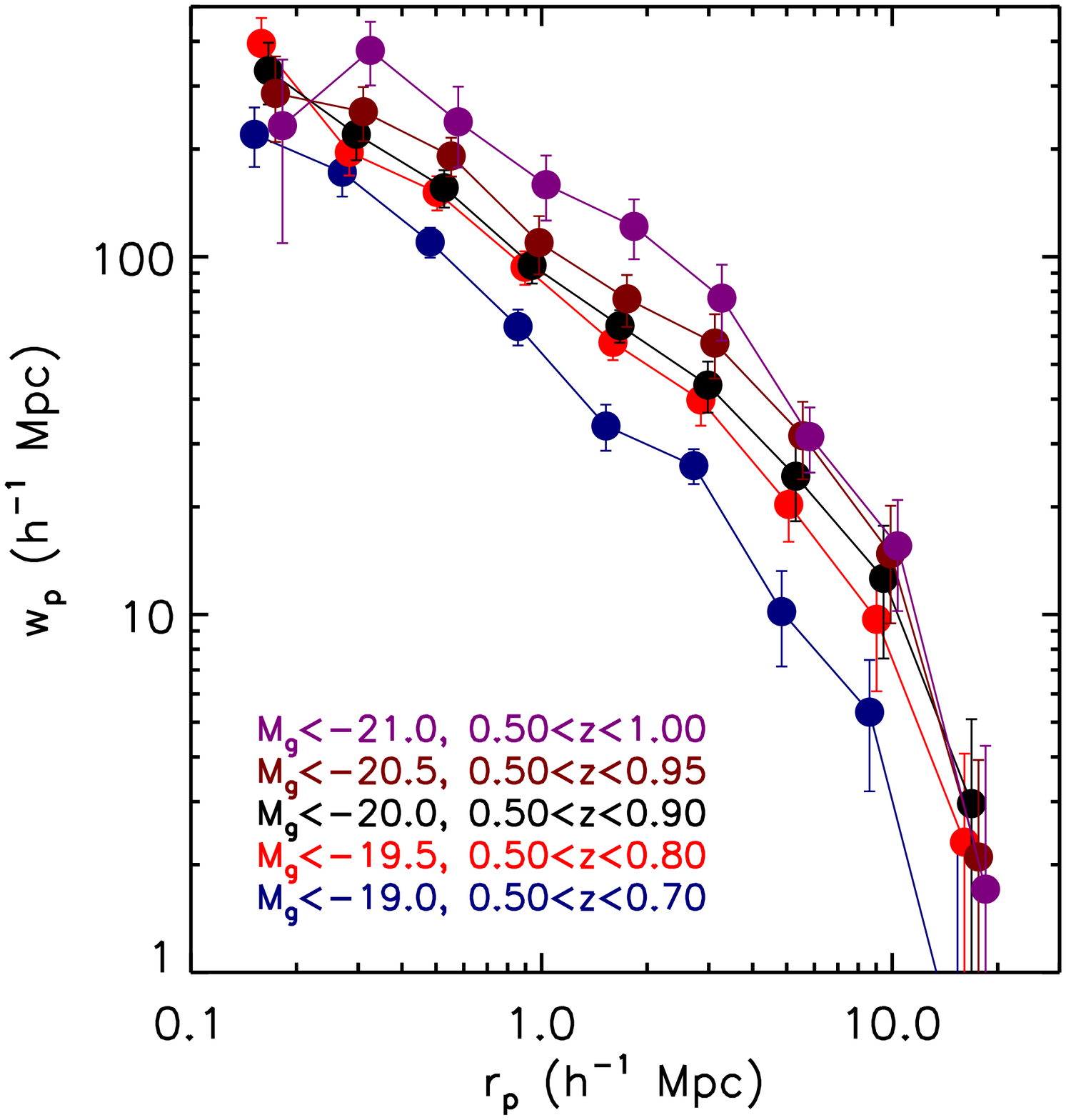} 
 	\caption{The projected correlation functions for the PRIMUS volume-limited threshold samples in the redshift ranges 0.20$<z<$0.50 (left) and 0.50$<z<$1.00 (right).  In the low-$z$ figure, the intermediate bins ($M_g<-17.5$, -18.5, and -19.5) are omitted, for clarity.  A clear luminosity dependence is visible, at both low- and high-redshift. 
        }
 	\label{fig:Ldepwp}
\end{figure*}	
		
The composite luminosity-dependent projected correlation functions are shown in Figure~\ref{fig:Ldepwp}, at lower redshift (left panel) and higher redshift (right panel). 
At a given galaxy separation in the figures, one can clearly see a trend with luminosity 
across the redshift range. 
The faintest low-redshift sample probes the smallest volume, and clustering at separations of $r_p>4~{\rm Mpc}/h$ cannot be robustly measured in it. 


Finally, we have also tested by performing these clustering measurements while excluding the COSMOS field from both the correlation functions and error analysis, and this decreases the clustering amplitude by $\sim20\%$ and $\sim 25\%$ at low-$z$ and high-$z$, respectively, and decreases the errors by $\sim30\%$ at low- and high-$z$. 
This resembles the effect of the Sloan Great Wall on clustering measurements in the SDSS (Zehavi et al.\ 2011; Norberg et al.\ 2011).  
It also highlights the dangers of interpreting high-$z$ galaxy clustering with the COSMOS field alone, where the large volume does not compensate for the effects of cosmic variance (e.g., McCracken et al.\ 2007; Meneux et al.\ 2009; de la Torre et al.\ 2010).
These issues are further discussed in Sections~\ref{sec:bevol}, \ref{sec:altstats}, and \ref{sec:var}.

\subsection{Power-law Fits}\label{sec:PL}
 
\begin{figure}[h!]
   	\includegraphics[width=1.0\linewidth]{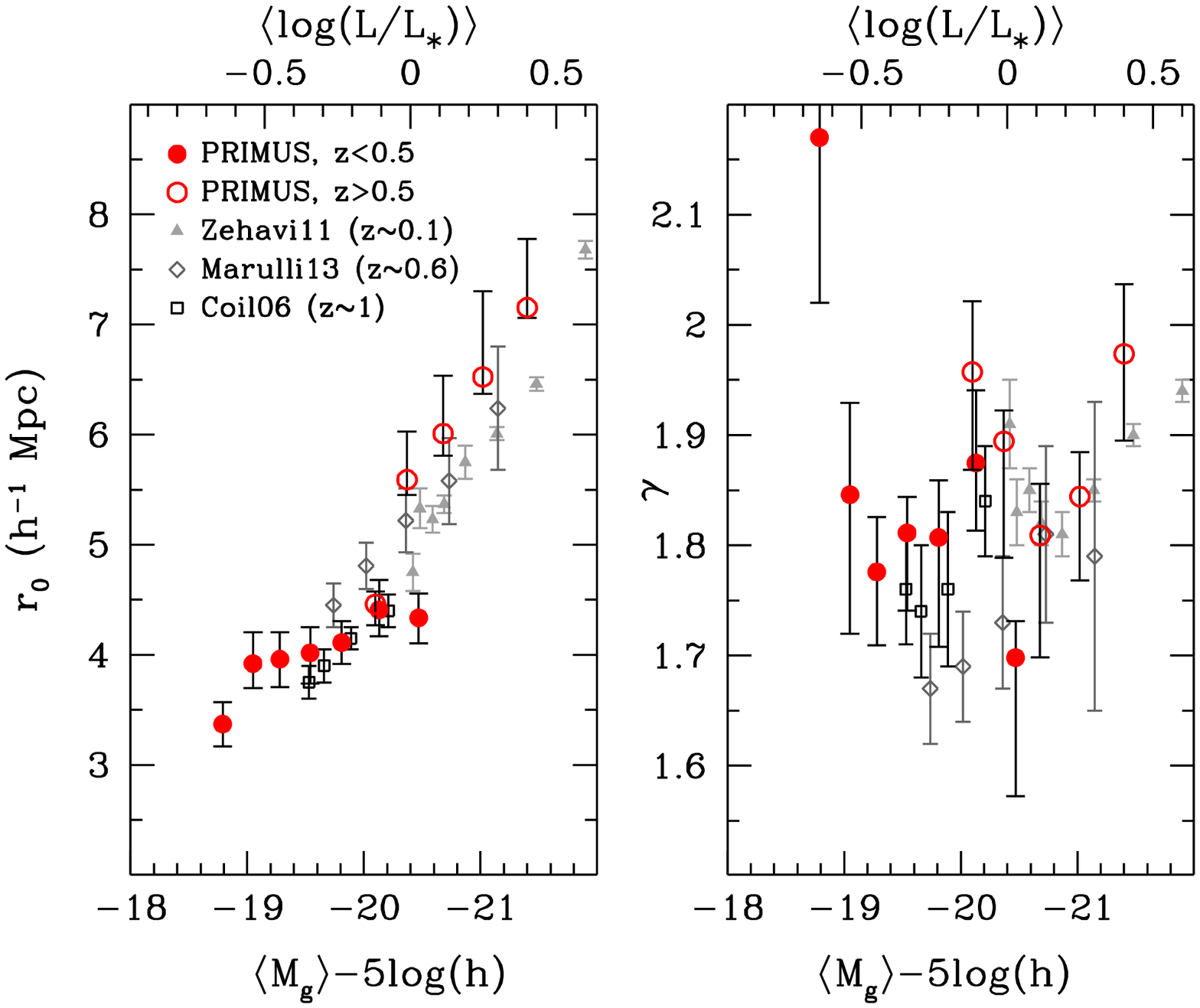} 
 	\caption{The clustering length ($r_0$) and slope ($\gamma$) for all the luminosity-threshold samples, estimated from power-law fits to the correlation functions.  
  The lower $x$-axis shows the mean $g$-band magnitude, while the upper one shows the $g$-band luminosity relative to $L_\ast$ of the luminosity function.  
  The filled red circles are the PRIMUS samples with 0.20$<z<$0.50 and the open red circles are those with 0.5$<z<$1.0. 
  The 1-$\sigma$ errors are estimated with the 16 and 84 percentiles of the jackknife subsamples.  
  Note that the high-$z$ $r_0$ results at $L\gtrsim L_\ast$ are affected by the large structure in the COSMOS field, which is in all but one of the subsamples; consequently, COSMOS raises the upper error bar. 
For comparison, we show the results of Coil et al.\ (2006b; DEEP2 survey; square points), Marulli et al.\ (2013; VIPERS; diamonds), and Zehavi et al.\ (2011; SDSS; triangles). 
        }
 	\label{fig:r0gamma}
\end{figure}

To quantify the luminosity-dependent clustering, we 
fit a power law to the correlation function $w_p(r_p)$ on scales of 0.5-$10~{\rm Mpc}/h$ to derive the parameters $r_0$ and $\gamma$ (defined with $\xi(r)=(r/r_0)^{-\gamma}$; see Eqn.~\ref{eq:powerlaw}), 
which characterize the clustering amplitude and slope. 
A power law fits all of our measured correlation functions well within the errors, except for the faintest bin, which we discuss further below.

The results are shown in Figure~\ref{fig:r0gamma} for the luminosity threshold samples, 
and these and the luminosity-binned results are listed in Tables~\ref{tab:r0gamma} and \ref{tab:r0gammaL}.  
We use both sets of samples as each has advantages: the luminosity bins are independent of each other, 
while the threshold samples have smaller errors and are useful for halo occupation distribution modeling and abundance matching 
because the thresholds translate into lower halo mass limits of integrals (see Sec.~\ref{sec:model}). 

\begin{table}
\caption{Clustering Results for Luminosity-Limited Samples: Power-law Fits and Bias}
 \centering
 \begin{tabular}{ l | c c c c c c }
   \hline
    $M_g^\mathrm{max}$ & $r_0$ (Mpc/$h$) & $\gamma$ & $b_{\rm gal}$ & $b_{\rm gal,HM}$ \\
   \hline
    & & $0.2<z<0.5$ & & \\
   \hline
   -17.0 & 3.37$\pm$0.20 & 2.17$\pm$0.15           & 0.90$\pm_{0.30}^{0.16}$ & 1.20$\pm$0.10 \\ 
   -17.5 & 3.92$\pm$0.25 & 1.93$\pm$0.10           & 1.29$\pm$0.10 & 1.23$\pm$0.10 \\ 
   -18.0 & 3.96$\pm$0.25 & 1.78$\pm$0.06           & 1.36$\pm$0.11 & 1.25$\pm$0.10 \\ 
   -18.5 & 4.02$\pm$0.26 & 1.81$\pm$0.05           & 1.26$\pm$0.11 & 1.28$\pm$0.10 \\ 
   -19.0 & 4.11$\pm$0.20 & 1.81$\pm$0.08           & 1.28$\pm$0.09 & 1.30$\pm$0.10 \\ 
   -19.5 & 4.41$\pm$0.26 & 1.87$\pm$0.06           & 1.34$\pm$0.09 & 1.34$\pm$0.10 \\ 
   -20.0 & 4.34$\pm$0.22 & 1.70$\pm_{0.13}^{0.03}$ & 1.38$\pm$0.17 & 1.38$\pm$0.10 \\ 
   \hline
    & & $0.5<z<1.0$ & & \\
   \hline
   -19.0 & 4.46$\pm$0.15 & 1.96$\pm$0.08                     & 1.15$\pm$0.18 & 1.40$\pm$0.20 \\ 
   -19.5 & 5.60$\pm_{0.14}^{0.44}$ & 1.89$\pm_{0.11}^{0.03}$ & 1.65$\pm$0.15 & 1.45$\pm$0.20 \\ 
   -20.0 & 6.01$\pm_{0.20}^{0.52}$ & 1.81$\pm$0.08           & 1.84$\pm$0.10 & 1.52$\pm$0.20 \\ 
   -20.5 & 6.52$\pm_{0.15}^{0.78}$ & 1.84$\pm$0.04           & 1.99$\pm_{0.12}^{0.18}$ & 1.61$\pm$0.25 \\ 
   -21.0 & 7.15$\pm_{0.10}^{0.63}$ & 1.97$\pm$0.07           & 2.09$\pm_{0.17}^{0.47}$ & 1.79$\pm$0.30 \\ 
   \hline
  \end{tabular}
 \begin{list}{}{}
    \setlength{\itemsep}{0pt}
    \item Results of power-law fits to $\xi(r)=(r/r_0)^{-\gamma}$ at 0.5-10~Mpc/$h$ and bias fits to $\xi(r)=b_g^2\xi_{mm}(r)$ on scales of 3-17~Mpc/$h$, for the luminosity threshold samples, which are described in Table~\ref{tab:samples}. 
    The halo-model (HM) bias values are listed in the far right column (see Sec.~\ref{sec:HMbias}).  
    Separate lower and upper error bars are quoted only in cases where they significantly differ, using the 16 and 84 percentiles of the jackknife subsamples. 
 \end{list}
 \label{tab:r0gamma}
\caption{Clustering Results for Luminosity-Binned Samples}
 \centering
 \begin{tabular}{ c c | c c c c }
   \hline
    $M_g^\mathrm{max}$ & $M_g^\mathrm{min}$ & $r_0$ (Mpc/$h$) & $\gamma$ & $b_{\rm gal}$ & $b_{\rm gal,HM}$ \\
   \hline
   -17.0 & -18.0 & 2.91$\pm_{0.48}^{0.66}$ & 2.33$\pm$0.31 & 0.79$\pm_{0.30}^{0.11}$ & 0.83$\pm$0.15 \\ 
   -18.0 & -19.0 & 3.67$\pm$0.32           & 1.94$\pm$0.09 & 1.05$\pm$0.11 & 0.87$\pm$0.15 \\ 
   -19.0 & -20.0 & 3.96$\pm$0.26           & 1.88$\pm$0.09 & 1.16$\pm$0.10 & 0.93$\pm$0.15 \\ 
   -20.0 & -21.0 & 4.20$\pm_{0.08}^{0.27}$ & 1.68$\pm$0.07 & 1.28$\pm_{0.08}^{0.24}$ & 1.02$\pm$0.15 \\ 
   \hline
   -19.0 & -20.0 & 4.08$\pm_{0.28}^{0.12}$ & 2.05$\pm$0.08 & 1.10$\pm_{0.11}^{0.25}$ & 1.08$\pm$0.20 \\ 
   -20.0 & -21.0 & 5.57$\pm_{0.26}^{0.43}$ & 1.79$\pm$0.07 & 1.77$\pm$0.16 & 1.30$\pm$0.25 \\ 
   -21.0 & -22.0 & 6.45$\pm_{0.21}^{0.63}$ & 1.82$\pm$0.07 & 2.00$\pm$0.38 & 1.60$\pm$0.30 \\ 
   \hline
  \end{tabular}
 \begin{list}{}{}
    \setlength{\itemsep}{0pt}
    \item Power-law fits for luminosity-binned samples, which are described in Table~\ref{tab:samples2}.  The columns are the same as in Table~\ref{tab:r0gamma}.  The upper set of results are for the $z<0.5$ samples, and the lower set of results are for the $z>0.5$ ones.
 \end{list}
 \label{tab:r0gammaL}
\end{table}

The correlation length $r_0$ increases rapidly with increasing luminosity (where $\langle M_g\rangle-5{\rm log}(h)$ is equivalent to mean luminosity, $\langle {\rm log}(L)\rangle$, and the trend as a function of threshold $L_{\rm min}$ is similar), from 3.4 to 7.2$~{\rm Mpc}/h$, implying that more luminous galaxies reside in more massive dark matter halos out to $z\sim1$. 
This trend is consistent with low- and high-$z$ studies of luminosity-dependent clustering (Coil et al.\ 2006b; Zehavi et al.\ 2011; Marulli et al.\ 2013; shown as the gray points in Fig.~\ref{fig:r0gamma}).  We discuss comparisons to these and other clustering and weak-lensing studies in the literature in Section~\ref{sec:literature}. 
In Figure~\ref{fig:r0gamma}, there is some indication of surprisingly strong high-$z$ clustering strength in the samples with $z\geq0.5$, but this difference is of weak statistical significance.  
We discuss this further below, in the context of evolving galaxy bias, in Section~\ref{sec:bevol}. 

The power-law slope of the correlation functions is within the range $\gamma\approx1.8$-$2.0$, as expected. 
In the faintest sample, we find a discrepancy with our overall trends in $r_0$ and $\gamma$, though these are partly due to poor power-law fits to the correlation functions, and are only of 2-$\sigma$ significance.  
Note that this sample covers the smallest volume of any of our samples, and might therefore be affected more by cosmic variance. 
In addition, when fits are done with fixed slope (e.g., $\gamma=1.9$), these discrepancies are partly alleviated (the faintest sample's $r_0$ increases to $\approx3.5~{\rm Mpc}/h$). 

\subsection{Galaxy Bias}

We now proceed to luminosity-dependent galaxy bias, which quantifies the degree to which galaxies are biased tracers of dark matter, and allows for another interpretation of large-scale clustering strength.  
Details about galaxy bias, and how it is interpreted with halo models of galaxy clustering, are described in Section~\ref{sec:HMbias}. 

We estimate the bias with respect to the nonlinear matter power spectrum of Smith et al.\ (2003; consistent within a few percent of the higher resolution $P(k)$ in Heitmann et al.\ 2010), 
and convert it to a projected correlation function with Eqn.~(\ref{eq:wp}) and $\pi_{\rm max}=80\,{\rm Mpc}/h$. 
This allows for a calculation of $b_{\rm gal}=\sqrt{w_{p,gg}(r_p,z)/w_{p,mm}(r_p,z)}$, following Eqn.~\ref{eq:bgal}. 

We perform this calculation over the range of $3<r_p<20~{\rm Mpc}/h$ and average over these scales, with $\langle z\rangle$ of the samples; using smaller scale $w_p$ ($1<r_p<10~{\rm Mpc}/h$) yields bias values that are similar or slightly lower (by $\approx10\%$). 
(We have also tested using the linear rather than nonlinear $P(k)$ for the $b_{\rm gal}$ calculation, and obtained bias values $\approx5\%$ higher.) 
The quoted 1-$\sigma$ errors are estimated from the distribution of the jackknife subsamples. 
Note that the bias factor depends on the amplitude of matter clustering $\sigma_8$, so that it is in fact $b_{\rm gal}\times(\sigma_8/0.8)$. 

\begin{figure}
  \includegraphics[width=\hsize]{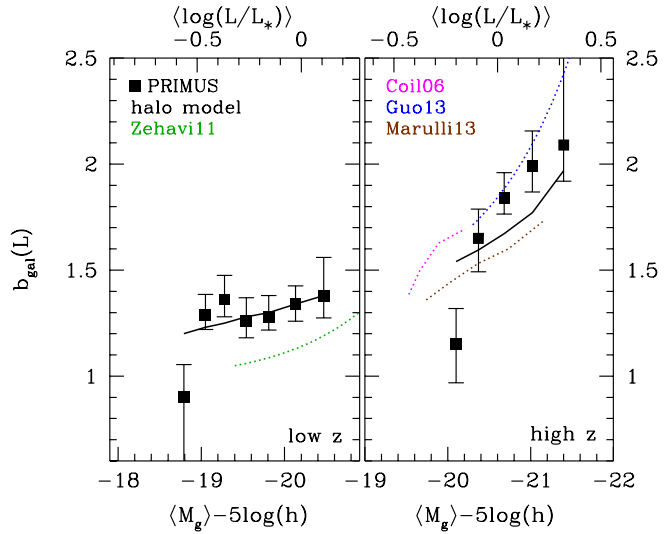} 
  \caption{Luminosity dependent bias comparison, with solid points for the luminosity threshold samples, and with errors indicated by the 16 and 84 percentiles of the jackknife subsamples. 
  Left panel: low-redshift results ($z<0.5$); right panel: high-redshift results ($z>0.5$). 
  Black solid lines are the predictions of a halo occupation model (see Sec.~\ref{sec:HMbias} for details). 
  Green dotted line shows the SDSS result for Zehavi et al.\ (2011) at $z\sim0.1$; the discrepancy with the other results is due to the redshift difference. 
  Magenta, blue, and brown dotted lines are for Coil et al.\ (2006b; DEEP2; we have accounted for their larger value of $\sigma_8$), 
  H.\ Guo et al.\ (2013; BOSS\protect\footnotemark[21]), 
  and Marulli et al.\ (2013; VIPERS), respectively. 
  Errors of the halo-model and literature biases are omitted, for clarity. 
  } 
 \label{fig:bLcomparison}
\end{figure}
\setcounter{footnote}{21}
\footnotetext{SDSS-III (Eisenstein et al.\ 2011) Baryon Oscillation Spectroscopic Survey (Dawson et al.\ 2013)}

The results are shown in Figure~\ref{fig:bLcomparison} and given in Tables~\ref{tab:r0gamma} and \ref{tab:r0gammaL}.  
Note that the $b_{\rm gal}(L)$ trends are qualitatively consistent with $r_0(L)$ in Figure~\ref{fig:r0gamma}.  
In agreement with previous studies (Norberg et al.\ 2001; Tegmark et al.\ 2004; Zehavi et al.\ 2005; Wang et al.\ 2007; Swanson et al.\ 2008a), we find that $b_{\rm gal}$ only weakly increases with luminosity at $L\leq L_\ast$, and rises more rapidly at brighter luminosities.  
In addition, the luminosity-dependent bias does not appear to evolve much from $z\sim0.2$ to $z\sim1$.\footnote{The bias does increase slightly over this redshift range, however.  Note that in the left panel of Fig.~\ref{fig:bLcomparison}, the Zehavi et al.\ (2011) result is at lower redshift than ours ($z\sim0.1$ versus $0.28<\langle z\rangle<0.39$) and the Coil et al.\ (2006b) result is at higher redshift ($z\sim1$ vs $0.60<\langle z\rangle<0.74$), so the apparent discrepancies with these results in the figure are an expected effect. Bias evolution is analyzed in more detail in Sec.~\ref{sec:bevol}.} 

Following previous work (e.g., Norberg et al.\ 2001; Zehavi et al.\ 2005), we fit a function to the luminosity dependent bias.  
Fitting to the results for the luminosity threshold samples in Figure~\ref{fig:bLcomparison}, we obtain the following:
\begin{equation}
 b_{\rm gal}(L) = 1.05 + 0.50 (L/L_\ast)^{1.10}
\end{equation}
This is a steeper function of luminosity than expected from the halo model prediction (see Sec.~\ref{sec:HMbias} below) and steeper than found by previous studies (with others obtaining a multiplicative factor of $\approx0.2$ rather than our $0.50\pm0.24$). 
This dependence appears to be driven by our high-$z$ clustering results.  However, our bias errors are somewhat large, and the uncertainties in these fitted parameters are large as well.

\section{Results: Color-dependent clustering}\label{sec:CdepCFs}

\subsection{Red versus Blue Galaxies}\label{sec:redblue}

\begin{figure}
   	\includegraphics[width=0.49\hsize]{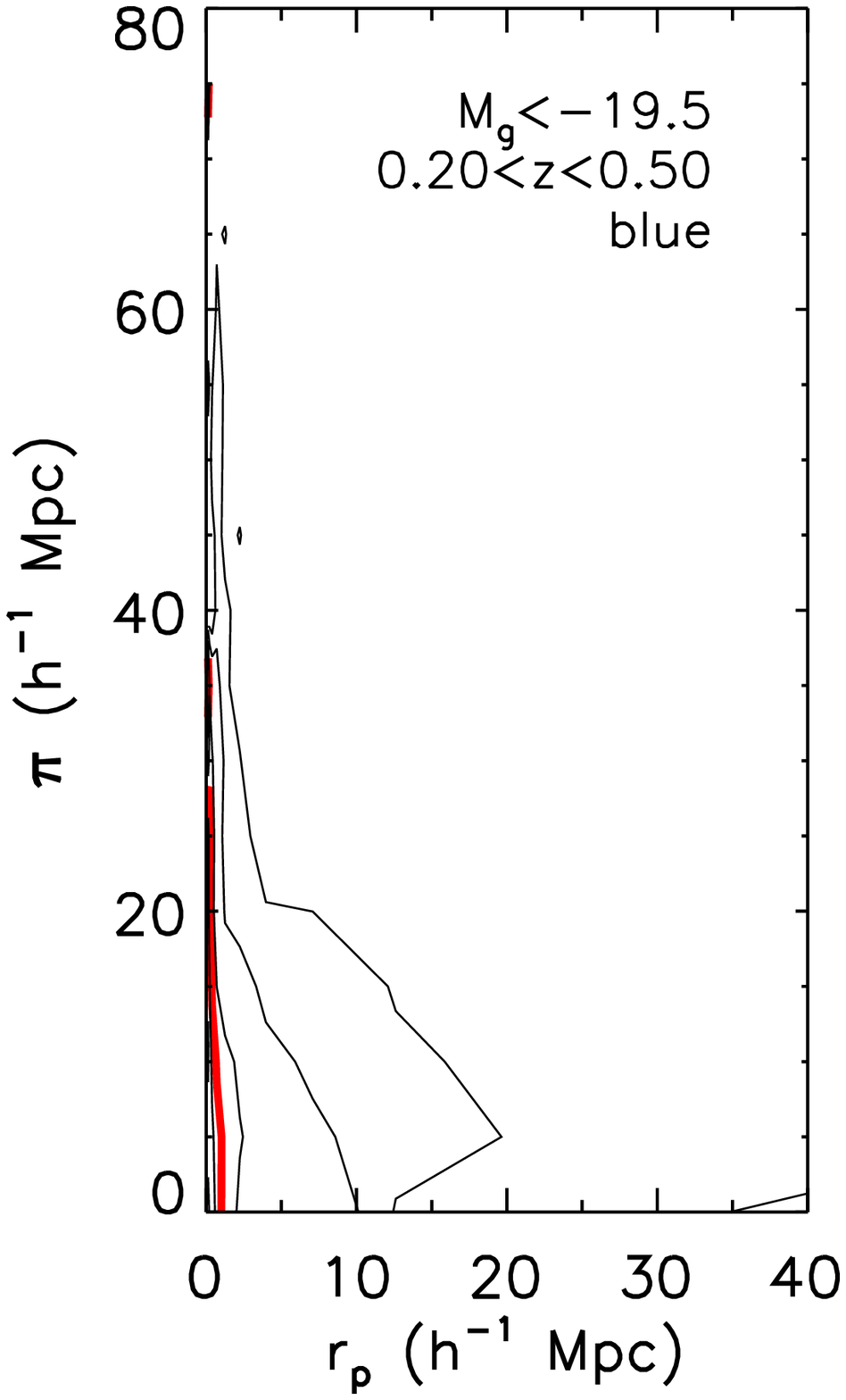} 
   	\includegraphics[width=0.49\hsize]{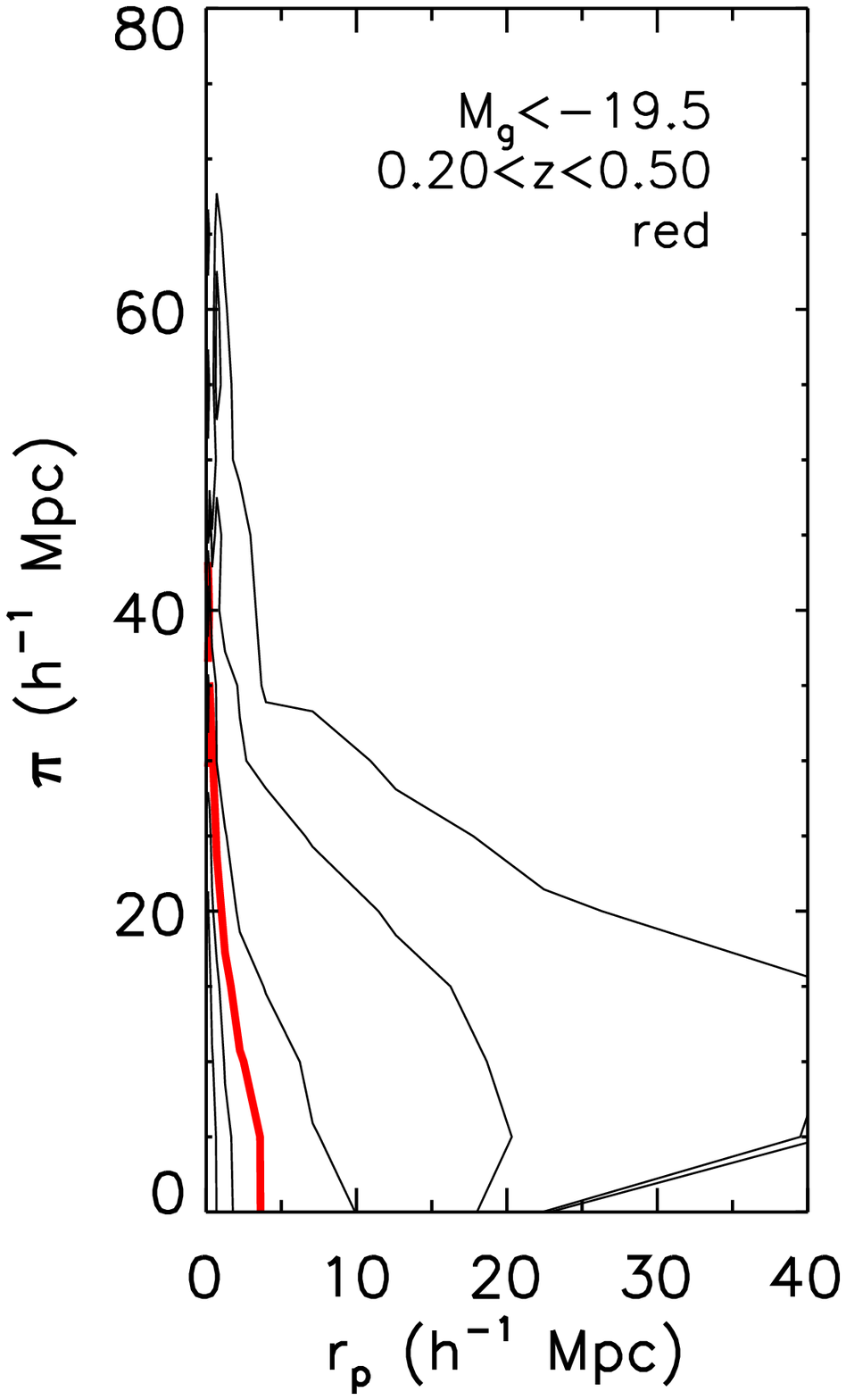} 
 	\caption{
          Two-dimensional correlation function $\xi(r_p,\pi)$ for $Q=4$ blue (left) and red (right) galaxies in the volume-limited threshold sample with $M_{\mathrm{g}}<-19.5$ and $0.20<z<0.50$.  
	  The contour levels are 0.2, 0.5, 1.0 (thick red line), 2.0, and 5.0. 
        } 
 	\label{fig:xirppiredblue}
\end{figure}

We first examine the clustering of red and blue galaxies as a function of luminosity, split according to the division in the color-magnitude distribution (Fig.~\ref{fig:color_samples}). 
We begin with the redshift-space correlation functions, shown in Figure~\ref{fig:xirppiredblue} for galaxies with $M_g<-19.5$. 
As seen for the full sample (Fig.~\ref{fig:xirppi}), the galaxies exhibit FOG elongations at small separations, due to virial motions within halos, as well as coherent infall of the halos themselves at large scales.  
However, red galaxies are clearly more strongly clustered than blue ones, at all scales. 
This is an expected result, as blue galaxies are better tracers of the ``field" (rather than dense environments) and are less biased. 
Nonetheless, it is interesting that we see statistically significant FOG distortions for blue galaxies, which is evidence for some blue galaxies residing in group and cluster environments (see also Coil et al.\ 2006a; Skibba 2009; Zehavi et al.\ 2011). From tests with mock catalogs, however, we find that the effect is slightly enhanced due to redshift uncertainties (i.e., small FOG appear larger). 


\begin{figure}
   	\includegraphics[width=\hsize]{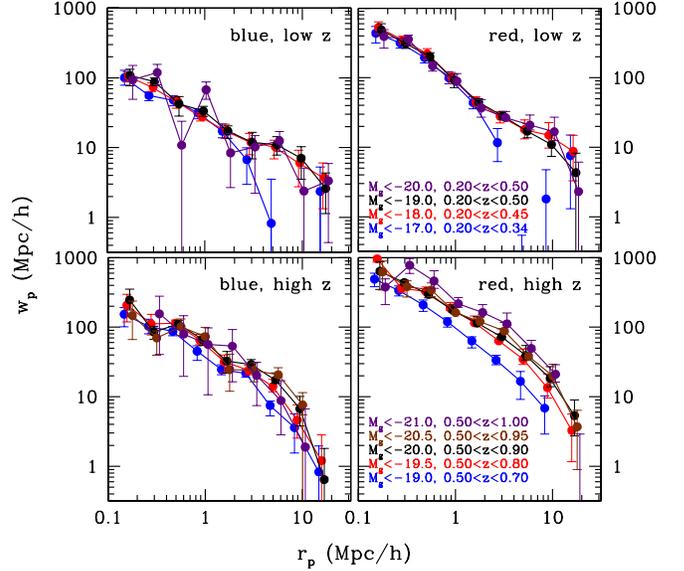} 
 	\caption{Projected correlation functions for blue and red galaxies (left and right panels), for $L$-threshold samples at low and high redshift (upper and lower panels). }
 	\label{fig:wpredblue}
\end{figure}

Next, we present the projected correlation functions of red and blue galaxies in luminosity bins of the low- and high-$z$ volume-limited catalogs in Figure~\ref{fig:wpredblue}. 
%
As seen for the full galaxy sample, a luminosity dependence of the correlation functions is visible, for both blue and red galaxies and at low and high redshift, though the uncertainties are somewhat larger due to the smaller sample sizes.  
Red galaxies are clearly more strongly clustered than blue ones at a given luminosity, as the red galaxy correlation functions (right panels) have amplitudes systematically larger by $\approx3$-$5\times$ at a given separation. 
This result shows that red sequence galaxies tend to reside in more massive dark matter halos than blue ones, a trend already in place at $z\sim1$ (Coil et al.\ 2008; Coupon et al.\ 2012). 
Note also that the red galaxies have a more pronounced one-halo term, especially at low $z$, which is an indication of a relatively large satellite fraction vis-\'{a}-vis blue galaxies. 
(The `one-halo term' refers to pairs of galaxies in a single halo, and dominates at small separations, $r_p<1$-$2~{\rm Mpc}/h$.  The `two-halo term' refers to galaxies in separate halos, and dominates at larger scales, in the linear regime.)

\begin{figure}
   	\includegraphics[width=1.0\hsize]{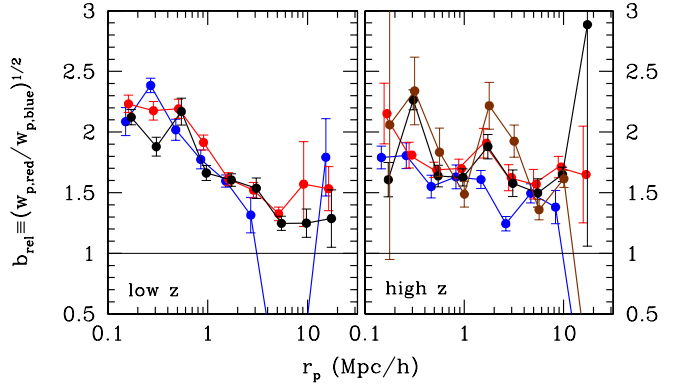} 
 	\caption{Relative bias of red to blue galaxies (at a given luminosity threshold), measured with the square root of the ratio of the projected correlation functions, $w_p(r_p)$. The color scheme is the same as in the upper and lower panels of Fig.~\ref{fig:wpredblue}. 
        Error bars indicate the rms of the jackknife subsamples. 
        The brightest samples (with poorest number statistics) are omitted, for clarity. 
        Red galaxies are clearly more strongly clustered than blue ones, and the effect is more pronounced at smaller projected separations. 
        }
 	\label{fig:brelredblue}
\end{figure}

We show this in Figure~\ref{fig:brelredblue}, with the relative bias of red and blue galaxies, 
$b_{\rm rel}\equiv[w_p(r_p|L)_{\rm red}/w_p(r_p|L)_{\rm blue}]^{1/2}$. 
The error bars are estimated from the rms of the $b_{\rm rel}$ of the jackknife subsamples. 
%
We find that $b_{\rm rel}>1$ at all separations, and we detect a significantly higher relative bias at small scales ($r_p<1~{\rm Mpc}/h$), consistent with Coil et al.\ (2008) and Meneux et al.\ (2006). 
There may be a weak luminosity dependence at $z>0.5$ as well, such that brighter galaxies have a slightly higher relative bias, consistent with an analysis of the color-density relation in Cucciati et al.\ (2006), though the trend is not detected in other clustering studies.  
These results suggests that the satellite fractions of red versus blue galaxies weakly depend on luminosity, though it is a small effect (see also Berlind et al.\ 2005; Skibba \& Sheth 2009). 
In any case, the scale dependence here, such that $b_{\rm rel}$ increases at smaller separations, confirms that the satellite fraction depends strongly on color, and is larger for red galaxies. 

\begin{figure}
   	\includegraphics[width=1.0\hsize]{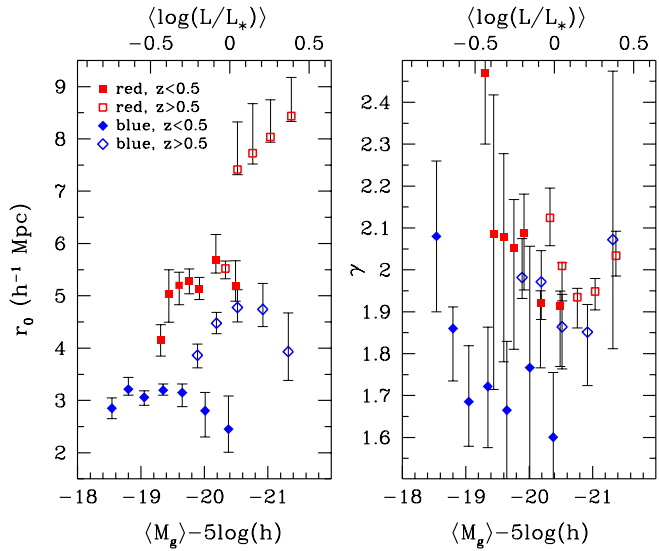} 
 	\caption{Clustering length ($r_0$, left) and slope ($\gamma$, right) for red and blue galaxies (indicated by red square and blue diamond points), as a function of luminosity.  
        Solid and open points are results at low and high redshift, respectively. 
        The DEEP2 and SDSS results of Coil et al.\ (2008) and Zehavi et al.\ (2011; not shown) are similar, though the latter have slightly higher red galaxy $r_0$ at $L\ll L_\ast$. }
 	\label{fig:r0redblue}
\end{figure}

We also perform power-law fits to the correlation functions, as in Sec.~\ref{sec:PL}.  
The resulting clustering lengths and slopes are shown in Figure~\ref{fig:r0redblue} and listed in Table~\ref{tab:r0gammaC}. 
At low and high redshift, especially for red galaxies, $r_0$ increases with increasing luminosity, and the trend steepens at $L\geq L_\ast$. 
However, for a given luminosity range, the luminosity dependence is much stronger for the red sequence galaxies than for the blue ones; the blue galaxies are consistent with a constant clustering length. 

The slope $\gamma$ has a wider dispersion than for the full catalogs, probably due to poorer number statistics.  
Nonetheless, there is a weak anti-correlation between $\gamma$ and luminosity, in particular with a slightly steeper slope for red galaxies.  This weak trend is seen in the DEEP2 and SDSS results of Coil et al.\ (2008) and Zehavi et al.\ (2011) as well. 

\begin{table*}
\caption{Clustering Results for Blue and Red Samples}
 \centering
 \begin{tabular}{ l | c c c | c c c }
   \hline
    $M_g^\mathrm{max}$ & blue $r_0$ & blue $\gamma$ & blue $b_{\rm gal}$ & red $r_0$ & red $\gamma$ & red $b_{\rm gal}$ \\
   \hline
   -17.0 & 2.85$\pm$0.20           & 2.08$\pm$0.15 & 0.78$\pm$0.16           & 4.15$\pm$0.30 & 2.47$\pm$0.15                     & 1.23$\pm$0.28 \\
   -17.5 & 3.21$\pm_{0.11}^{0.23}$ & 1.86$\pm$0.09 & 0.99$\pm$0.07           & 5.04$\pm$0.50 & 2.09$\pm$0.35                     & 1.39$\pm$0.14 \\
   -18.0 & 3.06$\pm$0.14           & 1.69$\pm$0.12 & 1.14$\pm$0.07           & 5.20$\pm$0.31 & 2.08$\pm$0.25                     & 1.46$\pm$0.12 \\
   -18.5 & 3.20$\pm$0.11           & 1.72$\pm$0.14 & 1.09$\pm$0.10           & 5.28$\pm$0.24 & 2.05$\pm_{0.24}^{0.11}$           & 1.48$\pm$0.11 \\
   -19.0 & 3.15$\pm_{0.27}^{0.16}$ & 1.67$\pm$0.17 & 1.13$\pm$0.16           & 5.14$\pm$0.21 & 2.09$\pm$0.11                     & 1.46$\pm$0.10 \\
   -19.5 & 2.80$\pm_{0.50}^{0.35}$ & 1.77$\pm$0.29 & 1.08$\pm$0.15           & 5.69$\pm_{0.25}^{0.49}$ & 1.92$\pm_{0.15}^{0.03}$ & 1.56$\pm$0.10 \\
   -20.0 & 2.45$\pm_{0.44}^{0.63}$ & 1.60$\pm$0.15 & 1.05$\pm$0.20           & 5.19$\pm_{0.29}^{0.48}$ & 1.91$\pm_{0.14}^{0.04}$ & 1.49$\pm$0.27 \\
   \hline
   -19.0 & 3.86$\pm$0.23           & 1.98$\pm$0.07 & 1.04$\pm_{0.15}^{0.25}$ & 5.52$\pm$0.17 & 2.12$\pm$0.07                     & 1.56$\pm$0.13 \\
   -19.5 & 4.48$\pm$0.20           & 1.97$\pm$0.08 & 1.26$\pm_{0.23}^{0.11}$ & 7.41$\pm_{0.08}^{0.91}$ & 2.01$\pm_{0.09}^{0.01}$ & 2.05$\pm_{0.11}^{0.34}$ \\
   -20.0 & 4.78$\pm$0.31           & 1.86$\pm$0.09 & 1.35$\pm$0.16           & 7.73$\pm_{0.21}^{0.95}$ & 1.93$\pm$0.05           & 2.35$\pm_{0.08}^{0.40}$ \\
   -20.5 & 4.74$\pm_{0.33}^{0.50}$ & 1.85$\pm$0.10 & 1.62$\pm$0.29           & 8.04$\pm_{0.06}^{0.71}$ & 1.95$\pm$0.04           & 2.39$\pm_{0.08}^{0.34}$ \\
   -21.0 & 3.93$\pm_{0.55}^{0.74}$ & 2.07$\pm$0.33 & 1.13$\pm$0.36           & 8.43$\pm_{0.04}^{0.74}$ & 2.03$\pm$0.05           & 2.92$\pm_{0.14}^{0.37}$ \\
   \hline
  \end{tabular}
 \begin{list}{}{}
    \setlength{\itemsep}{0pt}
    \item Power-law fits and bias results for the clustering of red and blue galaxies in luminosity-threshold samples (described in Table~\ref{tab:samples3}). 
    Separate lower and upper error bars are quoted only in cases where they significantly differ, using the 16 and 84 percentiles of the jackknife subsamples. 
 \end{list}
 \label{tab:r0gammaC}
\end{table*}

\begin{figure}
  \includegraphics[width=1.0\hsize]{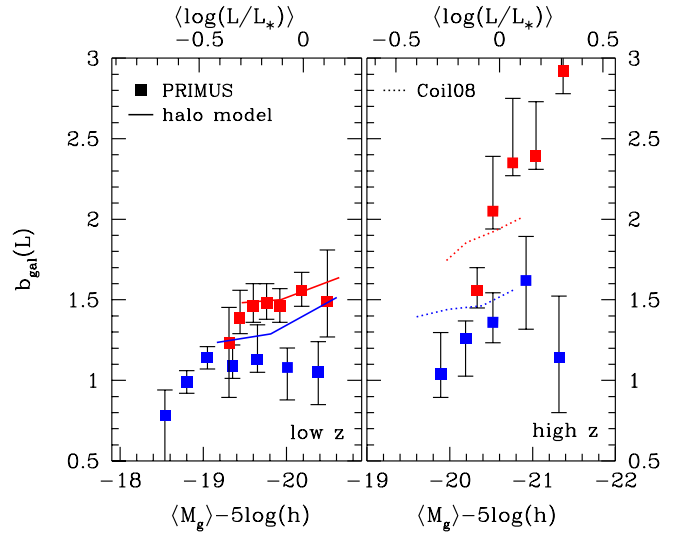} 
 	\caption{ 
        Luminosity dependent galaxy bias for blue and red galaxies, at low redshift ($z<0.5$, left panel) and high redshift ($z>0.5$, right panel). 
        Square points indicate the PRIMUS results, solid lines are the halo-model calculations (see Sec.~\ref{sec:HMbias} for details), and dotted lines are the Coil et al.\ (2008) DEEP2 results. }
  \label{fig:bgal}
\end{figure}

Finally, we present the luminosity-dependent bias of red and blue galaxies in Figure~\ref{fig:bgal} and Table~\ref{tab:r0gammaC}. 
Red galaxies are significantly more strongly biased than blue ones, at any given luminosity or redshift, qualitatively consistent with the $r_0$ trends in the previous figure. 
Moreover, for red galaxies, the bias steepens at $L>L_\ast$, consistent with the results in Figure~\ref{fig:bLcomparison}.  
For blue galaxies, the bias slightly decreases at bright luminosities, which may partly explain the discrepancy between the measurement and the halo-model prediction, although the model may not entirely reflect the division between red and blue galaxy clustering (Hearin \& Watson 2013). 

\subsection{Finer Color Bins}\label{sec:finecol}

To analyze the color dependence of clustering in more detail, we now study the correlation functions for narrow bins in color, using the color-magnitude cuts described in Section~\ref{sec:colcats}. 
As stated there, the clustering measurements do not depend significantly on the assumed color-magnitude cuts, their redshift evolution, or on the field-to-field variation of the color-magnitude distributions. 
The color bins are selected from a volume-limited sample with $M_g\leq19$ and $0.2<z<0.8$, which covers a wide dynamic range in magnitude and color.  
Each of the red and blue samples consist of $\sim6100$-6300 galaxies, while the sample of green valley galaxies is smaller (see Table~\ref{tab:samples4} for details).

\begin{figure}
   	\includegraphics[width=1.0\hsize]{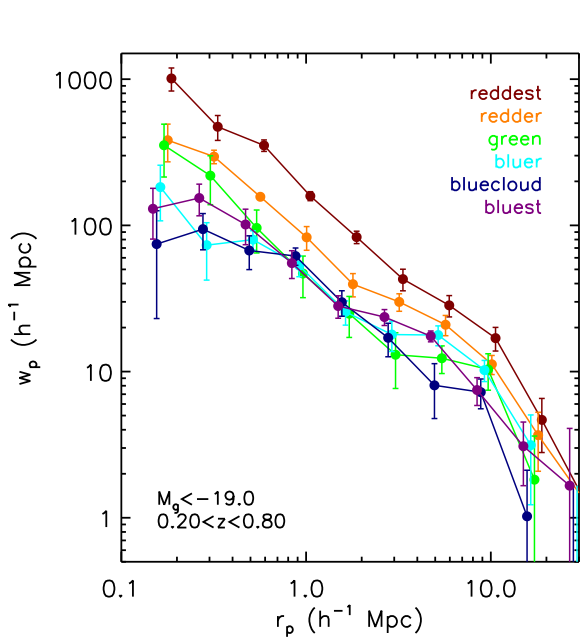} 
 	\caption{
        Projected correlation functions for finer color bin samples (described in Table~\ref{tab:samples4}): two (three) bins for the red (blue) sequence, and an intermediate `green valley' bin. 
        The points are slightly offset in ${\rm log}(r_p)$, for clarity.} 
 	\label{fig:wpfinecol}
\end{figure}

The projected correlation functions using the finer color bins are shown in Figure~\ref{fig:wpfinecol}. 
One can see a clear color dependence of the clustering amplitude, especially within the red sequence. 
Some have found that the clustering amplitude of green valley galaxies lies intermediately between that of blue and red galaxies (Coil et al.\ 2008; Zehavi et al.\ 2011; Krause et al.\ 2013), though our uncertainties are too large to determine this. 

The blue sequence galaxies do not have distinctly separated correlation functions, unlike some other studies (Coil et al.\ 2008; Zehavi et al.\ 2011; cf.\ sSFR-dependent clustering in Mostek et al.\ 2013). 
However, the samples used here are relatively faint and distinguishing between these low bias values ($b_{\rm gal}\approx0.9$-$1.3$) is not possible given the number statistics constraints. 
Interestingly, the bluest galaxies appear to have slightly stronger clustering than other blue galaxies and green valley galaxies, contrary to the expected monotonic color dependence of clustering strength.  This effect persists when the blue sequence is divided into two rather than three samples, and when other redshift and luminosity limits are used.  
However, 
the stronger clustering signal is of weak statistical significance (only $\sim$1-2$\sigma$).


\begin{table}
\caption{Clustering Results for Finer Color Samples}
 \centering
 \begin{tabular}{ l | c c c c }
   \hline
    color bin & $r_0$ (Mpc/$h$) & $\gamma$ & $b_{\rm gal}$ & $b_{\rm gal,HM}$ \\
   \hline
   bluest   & 4.24$\pm$0.26           & 1.92$\pm$0.08           & 1.23$\pm_{0.21}^{0.11}$ & 1.17$\pm$0.10 \\
   bluecloud & 3.87$\pm$0.38          & 1.98$\pm$0.12           & 0.97$\pm$0.16           & 1.16$\pm$0.10 \\
   bluer    & 3.99$\pm$0.31           & 1.74$\pm$0.09           & 1.28$\pm_{0.10}^{0.19}$ & 1.31$\pm$0.10 \\
   green    & 3.96$\pm_{0.19}^{0.35}$ & 1.97$\pm_{0.23}^{0.08}$ & 1.11$\pm_{0.26}^{0.19}$ & 1.42$\pm$0.15 \\
   redder   & 5.23$\pm_{0.12}^{0.33}$ & 1.97$\pm_{0.10}^{0.03}$ & 1.49$\pm_{0.08}^{0.16}$ & 1.45$\pm$0.15 \\
   reddest  & 6.18$\pm_{0.13}^{0.49}$ & 2.18$\pm_{0.14}^{0.06}$ & 1.69$\pm_{0.12}^{0.23}$ & 1.46$\pm$0.15 \\
   \hline
  \end{tabular}
 \begin{list}{}{}
    \setlength{\itemsep}{0pt}
    \item Power-law fits and bias results for finer color bins, from the sample with $M_g<-19$ and $0.2<z<0.8$ (described in Table~\ref{tab:samples4}). 
 \end{list}
 \label{tab:Cdepresults}
\end{table}

\begin{figure}
 \includegraphics[width=\hsize]{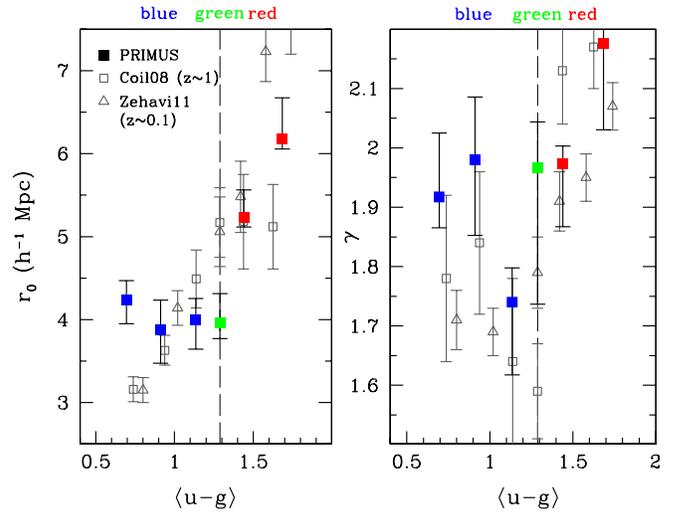} 
 \caption{Clustering length ($r_0$, left) and slope ($\gamma$, right) as a function of $u-g$ color. 
          For comparison, we show the results of Coil et al.\ (2008; open square points) and Zehavi et al.\ (2011; open triangles).
         }
 \label{fig:r0color}
\end{figure}

As in previous sections, we perform power-law fits to the correlation functions, and we present the resulting clustering amplitude and slope, $r_0$ and $\gamma$, in Table~\ref{tab:Cdepresults} and Figure~\ref{fig:r0color}.  
Except for the bluest galaxies, the values are approximately consistent with those of Coil et al.\ (2008) and Zehavi et al.\ (2011). 
The clustering strength clearly varies more within the red sequence than the blue cloud.  In addition, there is a weak trend such that the clustering power-laws are slightly steeper for redder galaxies, similar to the trend for red and blue galaxies (Fig.~\ref{fig:r0redblue}). 

\begin{figure}
 \includegraphics[width=\hsize]{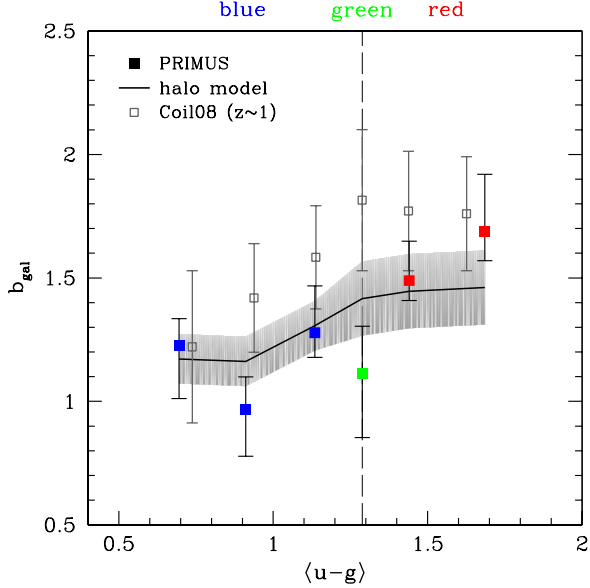} 
 \caption{PRIMUS (filled points) color-dependent bias, for the finer color binned samples.  
          For comparison, the results from the halo model-based mock catalog (black solid line) and Coil et al.\ (2008; open points) are shown.  
          As before, the Coil et al.\ results are shifted to account for the different values of $\sigma_8$; but note that their results are at $z\sim1$, while ours have a mean redshift of $\langle z\rangle\approx0.5$.  
          } 
 \label{fig:bCcomparison}
\end{figure}

Next, in Figure~\ref{fig:bCcomparison}, we present the color dependent galaxy bias.  
The blue and green valley galaxies have relatively low bias values, while the red sequence galaxies have larger bias, as expected from Figure~\ref{fig:bgal} and the color dependence of the correlation functions (Fig.~\ref{fig:wpfinecol}). 
The color dependence of bias very similar, but not exactly the same, as the the $r_0$ trends in Figure~\ref{fig:r0color} because the inferred clustering lengths also depend on the slope $\gamma$. 
The color trend is qualitatively consistent with the halo-model prediction (see Sec.~\ref{sec:HMbias}) and with the DEEP2 results from Coil et al.\ (2008), though note that the latter is at higher redshift ($z\sim1$ versus $\langle z\rangle\approx 0.5$). 
These comparisons are discussed further in Sections~\ref{sec:mock} and \ref{sec:literature}. 
Finally, in Section~\ref{sec:altstats}, 
we present and discuss color \textit{marked} correlation functions, which previously have only been measured in the SDSS (Skibba \& Sheth 2009).  

\section{Galaxy Bias and a Model of Color-Dependent Clustering}\label{sec:model}

The purpose of this section is to interpret the observed galaxy clustering trends in the context of the underlying dark matter distribution and the large-scale structure of DM halos. 
In Section~\ref{sec:HMbias}, we describe galaxy bias, and we compare simple halo-model predictions to PRIMUS measurements. 
In Section~\ref{sec:mock}, we use a halo-model description of galaxy colors and clustering to construct a mock galaxy catalog, which we compare to measurements of color-dependent clustering.  
More detailed models and mock catalogs, and further tests of them, will be the focus of a subsequent paper. 

\subsection{Calculating and Modeling Galaxy Bias}\label{sec:HMbias}

Dark matter halos are biased tracers of the underlying distribution of dark matter.  
In the $\Lambda$CDM theory of hierarchical structure formation, the large-scale clustering of halos with respect to matter can be described with: 
\begin{equation}
  \xi_{hh}(r,m,z) \approx [b_{\rm halo}(m,z)]^2\xi_{mm}(r,z)
\end{equation}
(Mo \& White 1996; Sheth \& Lemson 1999), where the matter correlation function is obtained from the linear or nonlinear power spectrum (Efstathiou, Bond \& White 1992; Smith et al.\ 2003).  
The halo bias $b_{\rm halo}$ has been derived from models of the halo mass function and from analyses of numerical simulations (Sheth, Mo \& Tormen 2001; Tinker et al.\ 2010). 

In the halo model of \textit{galaxy} clustering, galaxy bias $b_{\rm gal}$ can then be inferred from the abundance and bias of halos, combined with the occupation distribution of galaxies in the halos: 
\begin{equation}
  b_{\rm gal} = \int_{m_{\rm min}}^{m_{\rm max}} dm \frac{dn(m,z)}{dm} b_{\rm halo}(m,z) \frac{\langle N_{\rm gal}|m\rangle}{\bar n_{\rm gal}}
 \label{eq:HMbgal}
\end{equation}
(Cooray \& Sheth 2002; Yang et al.\ 2003), where the integration limits are related to the (e.g. luminosity dependent) selection of the galaxies themselves, 
$dn/dm$ is the halo mass function, 
$\langle N|m\rangle$ is the mean of the halo occupation distribution (HOD) of galaxies (a sum of $N_{\rm cen}$ and $N_{\rm sat}$ central and satellite galaxies), and 
$\bar n_{\rm gal}$ is the mean galaxy number density: 
\begin{equation}
  \bar n_{\mathrm {gal}} = \int_{m_{\rm min}}^{m_{\rm max}} dm \, {dn(m,z)\over dm}\, 
    \Bigl[\langle N_{\mathrm {cen}}|m\rangle + \langle N_{\mathrm {sat}}|m\rangle\Bigr]
 \label{eq:ngal}
\end{equation}

In terms of the large-scale galaxy correlation function, which depends on galaxy luminosity $L$, 
galaxy bias can be described as 
\begin{equation}
 \xi_{gg}(r|L,z) = [b_{\rm gal}(L,z)]^2 \xi_{mm}(r,z)
 \label{eq:bgal}
\end{equation}
though in practice, the bias is scale-dependent as well (Smith et al.\ 2007; van den Bosch et al.\ 2013). 
From correlation function measurements $\xi(r)$ or $w_p(r_p)$, $b_{\rm gal}$ can then be estimated at a given separation or over a range of scales (see e.g., Zehavi et al.\ 2005, 2011; Coil et al.\ 2006b, 2008).  
These studies and the results in this paper have shown that brighter and redder galaxies are more biased ($b\gg1$) than fainter and bluer objects.

We presented measurements of luminosity-dependent bias in the PRIMUS survey in Figures~\ref{fig:bLcomparison} and \ref{fig:bgal}. 
For comparison with these results, we show halo-model calculations in those figures as well (solid lines). 
These provide constraints on the masses of DM halos that host the galaxies, and on the distribution of central and satellite galaxies (recall that halos are assumed to host a single central galaxy and a number of satellites).

For these calculations, we use Eqn.~(\ref{eq:HMbgal}) with the mean redshifts of the samples and their number densities, quoted in Table~\ref{tab:samples}. 
We assume a Tinker et al.\ (2008b) halo mass function, Sheth et al.\ (2001) halo bias, and the halo mass relations of Moster et al.\ (2010) for the integration limits. 

We use a model of the HOD similar to that in Skibba \& Sheth (2009) and Zheng et al.\ (2007), and we refer the reader to these papers for details. 
The Skibba \& Sheth (2009) model is constrained by the luminosity and color dependent clustering in the SDSS, and the luminosity function and color-magnitude distribution. 
(In particular, the model's clustering constraints are correlation functions and mark correlation   functions using $r$-band luminosity and $g-r$ color at $0.017<z<0.125$.) 
The mean central and satellite HODs are the following:
\begin{equation}
  \langle N_\mathrm{cen}|m\rangle \,=\, \frac{1}{2}\Biggl[1\,+\,\mathrm{erf}\Biggl(\frac{\mathrm{log}(m/m_\mathrm{min})}{\sigma_{\mathrm{log}m}}\Biggr)\Biggr]
\label{eq:NcenM}
\end{equation}
\begin{equation}
  \langle N_\mathrm{sat}|m\rangle = 
    \Biggl(\frac{m-m_0}{m_1^{ ' }}\Biggr)^\alpha .
\label{eq:NsatM}
\end{equation}
where, for simplicity, we assume that $m_1/m_{\rm min}\approx20$, which determines the critical mass above which halos typically host at least one satellite galaxy (within the selection limits), 
and that $\alpha\approx1$.  
In addition, we assume that the HOD parameters do not evolve with redshift (which is likely an oversimplification; see Zheng et al.\ 2007). 
We find that this model yields approximately consistent number densities, but note that the HOD has not yet been constrained by the measured correlation functions. Future work may reveal that a more complex model with an evolving HOD is warranted.  

We obtain clustering results from the model that are consistent with the PRIMUS measurements, especially at low redshift.  
At high redshift, the theory bias values are slightly lower, which may be due to either shortcomings of the model or to significant differences between the PRIMUS survey and the SDSS, on which the HOD model was based.
%
We find that the mean halo masses for these luminosity thresholds and redshift ranges vary between $10^{11-13.5}~M_\odot/h$ and the satellite fractions vary between 0.15-0.30, but more detailed analysis of the clustering constraints with halo occupation and other models is beyond the scope of this paper and is the focus of future work.

\subsection{Mock Catalog with Colors}\label{sec:mock}

To aid the interpretation of the color dependence of galaxy clustering in PRIMUS, we 
again apply the model of Skibba \& Sheth (2009), described in the previous section. 
With this model, in Muldrew et al.\ (2012) we populated Millennium Simulation (Springel et al.\ 2005) halos and constructed mock galaxy catalogs.  For details, we refer the reader to Muldrew et al.\ and Skibba et al.\ (2013).
Note that there are a few other approaches to halo models with colors in the literature (Simon et al.\ 2009; Masaki et al.\ 2013; Hearin \& Watson 2013). 

We apply a simple modification of the mock catalog, in order to apply it to this analysis of galaxy clustering in PRIMUS.  First, we rescale the luminosity function $p(L/L_\ast)$ so that it is approximately consistent with the $g$-band LF in PRIMUS.  In practice, this means slightly reducing the abundance of faint galaxies, as the faint-end slope is slightly shallower (see also Moustakas et al.\ 2013). 
Then, we rescale the mock catalog's color distribution as a function of luminosity, 
$p(u-g|L_g)$, to approximately match the PRIMUS distribution, 
which accounts for the wider red sequence and smaller red fraction in PRIMUS.  
Finally, we impose the same color-magnitude divisions that were used in Sec.~\ref{sec:colcats}. 
Some simplifying assumptions have been applied with regard to the halo mass dependence of the color distributions (see Skibba \& Sheth 2009)\footnote{An important assumption is that the color distribution at a fixed luminosity is independent of halo mass.  While there is some evidence in support of this assumption (see Skibba 2009), recent work has shown that in some regimes color distributions can depend on both luminosity and halo mass, especially for central galaxies (More et al.\ 2011; Hearin \& Watson 2013).}, and a more thorough analysis will be performed in a subsequent paper.

\begin{figure}
   	\includegraphics[width=1.0\hsize]{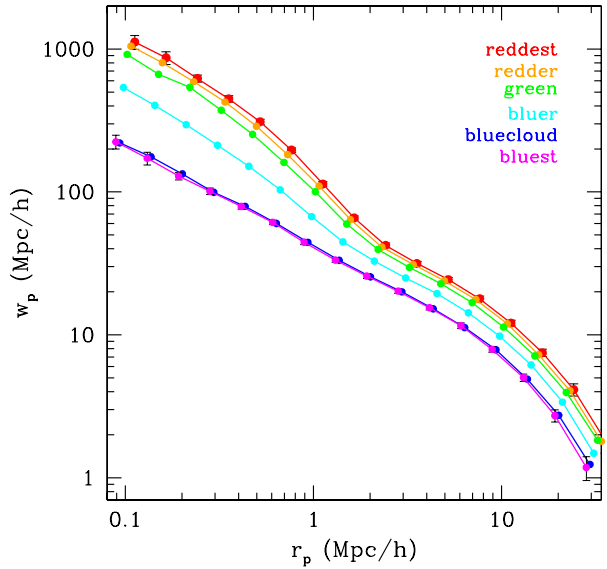} 
 	\caption{Projected correlation functions for finer color bin samples 
        in the mock catalog: two (three) bins for the red (blue) sequence, 
        and an intermediate green valley bin, as in Fig.~\ref{fig:wpfinecol}. 
        Error bars are shown for the reddest and bluest samples, and are small except at very small and very large separations. 
        The points are slightly offset in ${\rm log}(r_p)$, for clarity.} 
 	\label{fig:mockwpfinecol}
\end{figure}

We present the resulting color-dependent correlation functions in Figure~\ref{fig:mockwpfinecol}. 
We confirm the qualitative trends in Fig.~\ref{fig:wpfinecol}, where the reddest and bluest galaxies have a similar clustering strength as measured in the mock catalog. 
We see a wide range of color-dependent $w_p(r_p)$ at small scales ($r_p<1$-$2~{\rm Mpc}/h$), as in the data.  
These are mainly due to the higher fraction of satellite galaxies in the redder samples ($40\%$), resulting in an enhanced one-halo term (and steeper correlation functions). 
The reddest three samples have $w_p$ whose differences are less pronounced than in the data, however.  
This can also be seen in the color-dependent bias (Fig.~\ref{fig:bCcomparison}), in which the PRIMUS $b_{\rm gal}(u-g)$ appears to be slightly steeper than in the mock catalog.  
Simply rescaling the colors in the procedure above appears to be insufficient to reproduce the observed clustering, as the clustering amplitude of the `green' galaxies is too strong in the mock catalog. 

Another interesting result is that the `blue cloud' and `bluest' subsamples of the mock catalog have \textit{nearly identical} correlation functions.  
We find that these two subsamples have nearly identical mean halo masses ($1.4\times10^{13}M_\odot$), satellite fractions ($18\%$), and bias values ($1.17$). 
This partly occurs by construction, as in the Skibba \& Sheth (2009) model it is assumed that the color distribution at fixed luminosity is independent of halo mass; that is, $p(c|L,m)\approx p(c|L)$. 
The constant clustering of blue galaxies in PRIMUS appears to be consistent with this, though the discrepancy within the red sequence, such that the reddest galaxies are more strongly clustered, may indicate a regime in which this assumption breaks down. 
It is also interesting that in Zehavi et al.\ (2011) and Coil et al.\ (2008), a stronger color dependence of clustering strength and bias among blue galaxies is observed, compared to both PRIMUS and this mock catalog; considering that all of these analyses accounted for relevant incompletenesses, the origin of this different color dependence is not clear.

In any case, the largest discrepancies between the mock catalog and our PRIMUS results occur for the intermediate samples, especially for the blue cloud and green valley galaxies. 
Perhaps the clustering of these galaxies in PRIMUS is biased low, or perhaps cosmic variance is playing a role\footnote{However, we have attempted to account for this by using the 16 and 84 percentiles, rather than the variance of the jackknife subsamples, for the errors.}, but the discrepancies are of $\sim2$-$\sigma$ significance. 
We further discuss these results compared to results in the literature in Section~\ref{sec:literature}.

\section{Discussion}\label{sec:discussion}

We now briefly discuss our results in the context of the literature (Sec.~\ref{sec:literature}), galaxy clustering evolution (Secs.~\ref{sec:bevol} and \ref{sec:altstats}), galaxy evolution and star formation quenching (Sec.~\ref{sec:galevol}), and cosmic variance (Sec.~\ref{sec:var}).

\subsection{Comparisons to the Literature}\label{sec:literature}


A variety of related galaxy clustering analyses have been performed with low-redshift surveys (e.g., 2dF, SDSS, GAMA), intermediate-redshift surveys (BOSS, VIPERS), and high-redshift surveys (VVDS, DEEP2, zCOSMOS, CFHTLS\footnote{Canada-France-Hawaii Telescope Legacy Survey (Goranova et al.\ 2009).}, COMBO-17\footnote{Classifying Objects by Medium-Band Observations in 17 Filters (Wolf et al.\ 2004).}), and many of these have already been mentioned in this paper. 
Some of these, especially those with spectroscopic redshifts and real-space or projected correlation functions, allow for comparisons of trends and results, especially of the luminosity and color dependence of clustering length and slope ($r_0$ and $\gamma$) and galaxy bias.

Our measurements of $r_0$ and $\gamma$ (Figs.~\ref{fig:r0gamma}, \ref{fig:r0redblue}, and \ref{fig:r0color}) are generally consistent within their errors with $z\sim0.1$ measurements (Norberg et al.\ 2002; Zehavi et al.\ 2011) and $z\sim1$ measurements (Pollo et al.\ 2006; Meneux et al.\ 2006; Coil et al.\ 2006b, 2008). 
However, our clustering lengths for luminous and luminous red galaxies at $z>0.5$ are slightly large (by 1-2$\sigma$) compared to these studies, though when the COSMOS field is excluded from our measurements, the tension is minimal. 
Our luminosity-dependent results can be more directly compared to the intermediate-redshift analysis of Marulli et al.\ (2013), whose $r_0$ and $\gamma$ are consistent, but whose $z>0.5$ $b_{\rm gal}(L)$ values are slightly lower than ours (Fig.~\ref{fig:bLcomparison}) but are within the 1-$\sigma$ errors; our luminosity-dependent results are consistent with the recent analysis by Arnalte-Mur et al.\ (2013) as well. 
In addition, our results appear qualitatively consistent with those of more massive galaxies at similar redshifts (Wake et al.\ 2008; White et al.\ 2011; H.\ Guo et al.\ 2013). 

For our color-dependent results, we find good agreement with Coil et al.\ (2008) and Zehavi et al.\ (2011), except for the subsample of bluest galaxies, whose clustering length in Fig.~\ref{fig:r0color} is $\approx3\sigma$ higher than the SDSS and DEEP2 measurements. 
This is likely a statistical anomaly, as we have corrected for incompletenesses and the subsample's bias value is more consistent with the DEEP2 measurements and mock catalog (Fig.~\ref{fig:bCcomparison}). 

In our mock galaxy catalogs, on the other hand, we obtain nearly identical clustering lengths and biases for the two bluest subsamples (see Sec.~\ref{sec:mock}). 
However, for the reddest galaxies, the clustering in the mock is slightly low and the bias's slope appears too shallow.  This may be due to the fact that at a given luminosity, the color distribution of central galaxies depends on halo mass (More et al.\ 2011), an effect that may not be adequately addressed by the model.  
Nonetheless, the bias values are consistent within the errors. 

Finally, we note other related studies that have employed complementary methods to study the environmental dependence of galaxies as a function of luminosity and color, with qualitatively consistent trends.  
In particular, there have been a variety of relevant weak-lensing studies (Mandelbaum et al.\ 2006; Cacciato et al.\ 2009; Leauthaud et al.\ 2012), 
including of the DLS (Choi et al.\ 2012), which is one of the PRIMUS fields; 
studies of void probability functions (Conroy et al.\ 2005; Tinker et al.\ 2008a); 
three-point correlation functions and other clustering statistics (Croton et al.\ 2007; McBride et al.\ 2011a); 
counts-in-cells (Swanson et al.\ 2008a; Wolk et al.\ 2013); 
conditional luminosity functions (Cooray 2005; van den Bosch et al.\ 2007); 
and group and cluster catalogs (e.g., Weinmann et al.\ 2006; Hansen et al.\ 2009; Lin et al.\ 2014). 
In addition, the IMACS Cluster Building Survey (Oemler et al.\ 2013) uses the same telescope as PRIMUS, and has constraints on the star formation properties of galaxies as a function of environment (see Dressler et al.\ 2013), which could be compared to clustering as a function of SFR (Skibba et al., in prep.). 
%
Using different techniques, datasets, and different areas of the sky, these studies generally find that luminous and red galaxies are more strongly clustered than their fainter and bluer counterparts, at both small scales ($r<1~{\rm Mpc}/h$) and large scales ($r>1~{\rm Mpc}/h$), and at redshifts over the range $0<z<1$.  
Hansen et al.\ (2009) constrain color gradients within clusters, and Swanson et al.\ (2008a) analyze the luminosity-dependent bias of red and blue galaxies, but none of these study the environmental or clustering dependence using finer color bins, within the blue cloud or red sequence of the color-luminosity distribution.  
Therefore, our results in Section~\ref{sec:finecol} are particularly important and new by providing this.   

\subsection{Implications for Clustering Evolution}\label{sec:bevol}

We now discuss our results on the luminosity- and color-dependent clustering \textit{evolution}, and their implications for the relations between galaxies and the underlying matter density field, and for the growth of galaxies by mergers and star formation (Lackner et al.\ 2012; L'Huillier et al.\ 2012). 
These results provide new constraints, complementary to other results in the literature at lower and higher redshift (see Sec.~\ref{sec:literature} above). 

Our clustering measurements can be used to constrain departures from passive evolution (i.e., in the absence of merging, when the clustering follows the underlying density field), and potentially the amount of mergers between central and satellite galaxies. 
We now perform a simple test, by measuring the evolving galaxy bias, $b(z|L)$, compared to the trend predicted by passive evolution prediction (Fry 1996; Tegmark \& Peebles 1998).
Following Fry (1996), for a passively-evolving population, we assume that the bias evolution is given by:
\begin{equation}
  b_{\rm gal}(z) = \frac{(b(z_0)-1)D(z_0)}{D(z)} + 1,
 \label{eq:bz}
\end{equation}
where $D(z)$ is the linear growth factor for density perturbations. 
From this relation, the correlation function evolution for a passively-evolving population in the linear regime can be expressed as 
\begin{equation}
  \xi_{gg}(z) = [b_{\rm g}(z)D(z)]^2 \xi_{mm}(0) = [D(z)+(b_{\rm g}(z_0)-1)]^2 \xi_{mm}(0),
\end{equation}
where $\xi_{mm}(0)$ is the matter correlation function at $z=0$. 

Tojeiro \& Percival (2010) find that the number and luminosity density of bright `luminous red galaxies' (LRGs) are consistent with passive evolution, though our samples are dominated by fainter and bluer galaxies.  
Other studies also have evidence for small deviations from passive evolution over the redshift range $0.2<z<0.7$ (Conroy et al.\ 2007; White et al.\ 2007; Wake et al.\ 2008; H.\ Guo et al.\ 2013), and with halo models have argued that a significant fraction of satellites must either merge with central galaxies or be disrupted and become part of the intracluster light (ICL).

\begin{table*}
\caption{Number Density-Selected Catalogs}
 \centering
 \begin{tabular}{ l | c c c | c c c | c c c | c c c }
   \hline
    & & SDSS & & & PRIMUS & & & PRIMUS & & & DEEP2 & \\
    $\bar n_{\rm gal}$ & $M_r^\mathrm{max}$ & $z_{\rm min}$ & $z_{\rm max}$ & $M_g^\mathrm{max}$ & $z_{\rm min}$ & $z_{\rm max}$ & $M_g^\mathrm{max}$ & $z_{\rm min}$ & $z_{\rm max}$ & $M_B^\mathrm{max}$ & $z_{\rm min}$ & $z_{\rm max}$ \\
   \hline
   1.30 &  -19.3 & 0.02 & 0.064  &  -19.65 & 0.20 & 0.50  &  -19.25 & 0.50 & 0.80  &  -19.0 & 0.75 & 1.0 \\
   0.85 &  -19.75 & 0.02 & 0.07  &  -20.05 & 0.20 & 0.50  &  -19.85 & 0.50 & 0.85  &  -19.5 & 0.75 & 1.1 \\
   0.50 &  -20.2 & 0.02 & 0.10   &  -20.4  & 0.20 & 0.50  &  -20.35 & 0.50 & 0.90  &  -20.0 & 0.75 & 1.2 \\
   \hline
  \end{tabular}
 \begin{list}{}{}
    \setlength{\itemsep}{0pt}
    \item Catalogs selected at fixed number density, $\bar n_{\rm gal}$ (rather than at fixed luminosity thresholds) in units of $10^{-2}h^3\mathrm{Mpc}^{-3}$, used in Figure~\ref{fig:bz}.
    Each set of three columns describe the catalogs: $M_\lambda<M_\lambda^{\rm max}$ (or $L>L_{\rm min}$) and $z_{\rm min}<z<z_{\rm max}$.  The first set is for SDSS catalogs (similar to those of Zehavi et al.\ 2011), the next two are PRIMUS catalogs, and the last set is from DEEP2 results (Coil et al.\ 2006b). 
 \end{list}
 \label{tab:bgalzsamples}
\end{table*}

For the results in this paper, we have used samples with fixed luminosity thresholds $L_{\rm min}$ (or luminosity bins); however, such samples' number densities evolve, so that the high-$z$ galaxies are not necessarily progenitors of the low-$z$ ones (Tojeiro et al.\ 2012; Leja et al.\ 2013). 
In addition, over the mean redshift range of our samples, $L_\ast$ of the luminosity function evolves by more than a factor of $\approx2.2$ and $M_\ast$ of the halo mass function evolves by $\approx3$; the uncertainty in the redshift evolution of LFs also makes it difficult to use samples defined with $L/L_\ast(z)$. 
In light of this, we now use a common number density $\bar n_{\rm gal}$, which has recently been advocated and employed by these and other authors (H.\ Guo et al.\ 2013; Behroozi et al.\ 2013). 
This also allows for comparisons with SDSS ($z\sim0.1$) and DEEP2 ($z\sim1$) results, to widen the dynamic range of $b_{\rm gal}(z)$. 

\begin{figure}
 \includegraphics[width=0.85\hsize]{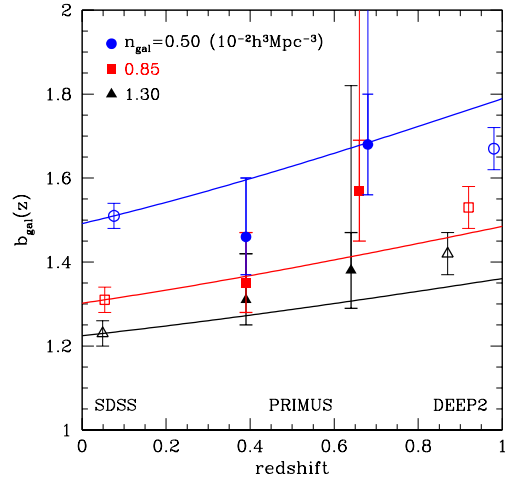} 
 \caption{ Galaxy bias evolution, $b_{\rm gal}(z)$, for galaxy samples selected at fixed number density, $\bar n_{\rm gal}$ (labeled in the upper left, in units of $10^{-2}h^3\mathrm{Mpc}^{-3}$). 
  The higher number density galaxies correspond to fainter galaxies; the lowest number density selection roughly corresponds to galaxies with $L>L_\ast(z)$. 
  The extended upper errors of the high-$z$ PRIMUS $b_\mathrm{gal}$ indicate the upper error bars when the COSMOS field is included. 
  For comparison, lines show passive evolution predictions relative to the SDSS bias factors, used as a low-$z$ benchmark (see text for details). }
 \label{fig:bz}
\end{figure}

We show the redshift-dependent bias in Figure~\ref{fig:bz}. 
Results for PRIMUS samples are shown, compared to SDSS samples 
and DEEP2 results (Coil et al.\ 2006b; accounting for the different value of $\sigma_8$). 
We perform the SDSS measurements with volume-limited samples from Data Release~7 (Abazajian et al.\ 2009), similar to the samples of Zehavi et al.\ (2011), in order to consistently compare to the PRIMUS and DEEP2 results; our bias values are slightly higher than those of Zehavi et al., likely due to the fitting being performed at slightly different scales. 
All of the samples are described in Table~\ref{tab:bgalzsamples}.  
For comparison, the prediction of passive evolution, using the relation in Eqn.~(\ref{eq:bz}), is also shown. 

Note that for the high-$z$ PRIMUS measurements, they are shown with the COSMOS field excluded, while the extended upper error bars indicate the upper errors of the higher bias values when the COSMOS field is included.  All three of these samples are affected by the large structure at $z\sim0.7$ in COSMOS. 
When $0.5<z<0.65$ is used (i.e., lower redshifts than the COSMOS structure) for these three samples and COSMOS is included, the bias results are very similar to the results in the figure; however, with a narrower redshift range, the errors are larger due to the reduced sample size. 
The bias for luminous galaxies (low number density) in the SDSS appears to be less affected by the Sloan Great Wall (see also Zehavi et al.\ 2011).

From the results in Figure~\ref{fig:bz}, there do not appear to be deviations from passive evolution over the redshift range $0<z<1$, though the uncertainties are too large to put significant constraints on $b(z)$. 
The fainter galaxies (higher number density) have smaller error bars, and their bias evolution is slightly more rapid than passive evolution, which is consistent with merger activity in these galaxy populations.  
Halo-model analyses of the full clustering measurements, including the small-scale one-halo term (see Wake et al.\ 2008), will provide further constraints and shed more light on this issue, as halo-subhalo mergers are expected to affect the small-scale clustering signal (Wetzel et al.\ 2009).  
The fact that $b(z)$ does not strongly depart from passive evolution, and that the clustering strength does not strongly evolve either (see Fig.~\ref{fig:r0gamma}), 
implies that the luminosity-halo mass relation does not strongly evolve from $z\sim1$ to $z\sim0$ (Conroy et al.\ 2006; Zheng et al.\ 2007). 

\subsection{Alternative Clustering Statistics}\label{sec:altstats}

We now present luminosity and color marked correlation functions at multiple redshifts, 
and obtain results consistent with those above. 

Mark clustering statistics of galaxies, and in particular, mark correlation functions, quantify how galaxy properties are correlated with the environment, as a function of scale. 
In essence, for each pair separation $r$, the statistic weights
each galaxy in a pair by its own attribute (e.g., luminosity, color,
etc., expressed in units of the mean across the population) and then
divides this weighted pair count by the unweighted one. 
Rather than splitting galaxy catalogs in luminosity and color bins, as is traditionally done, these statistics allow one to exploit the number statistics of the full catalog.  They are very sensitive to environmental correlations, and are useful for constraining halo models of galaxy clustering. 
For more details, we the refer the reader to Skibba et al.\ (2006), Skibba et al.\ (2013), and references therein.

The most commonly used mark statistic is the mark correlation function, which is defined as the ratio of the weighted/unweighted correlation function:
\begin{equation}
  M(r) \,\equiv\, \frac{1+W(r)}{1+\xi(r)}
 \label{markedXi}
\end{equation}
and the mark projected correlation function is similarly defined: 
$M(r_p)\equiv(1+W_p/r_p)/(1+w_p/r_p)$. 
If the weighted and unweighted clustering are significantly different at a particular separation $r$, then the mark is correlated (or anti-correlated) with the environment at that scale;
the degree to which they are different quantifies the strength of the correlation.

It can be difficult to compare mark correlations for different marks, as they may have different distributions, which could affect the clustering signals. 
In order to account for this, Skibba et al.\ (2013) introduced \textit{rank-ordered} mark correlations, which involve rank ordering the marks and using the rank itself as a weight. 
(In practice, we rank order and then match to a uniform distribution on [1,N].  In this way, all marks are scaled to the same distribution, so the mark correlation signal can be compared between
marks.)

\begin{figure}
 \includegraphics[width=\hsize]{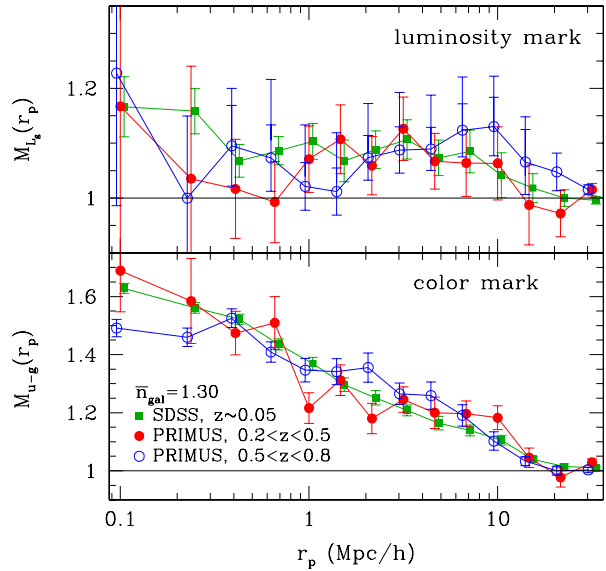} 
 \caption{Luminosity (upper panel) and color (lower panel) mark correlation functions for low and high-redshift PRIMUS samples ($z<0.5$ and $z>0.5$, solid and open circles), with the number density $\bar n_{\rm gal}=1.30~10^{-2}h^3\mathrm{Mpc}^{-3}$ (see Table~\ref{tab:bgalzsamples}). 
          The extended upper errors of the high-$z$ luminosity mark indicate the upper error bars when the COSMOS field is included. 
          SDSS results (green squares) are also shown, for comparison, and the points are slightly offset in ${\rm log}(r_p)$, for clarity. 
          There appears to be no evolution of the luminosity- and color-environment correlations between $z\sim0$ and $z\sim1$. 
         }
 \label{fig:MCF}
\end{figure}

We present rank-ordered luminosity and color mark correlation functions, for low- and high-redshift samples, in Figure~\ref{fig:MCF}. 
Results are shown for a fixed number density, $\bar n_{\rm gal}=1.30~10^{-2}h^3\mathrm{Mpc}^{-3}$, like the measurements in Sec.~\ref{sec:bevol} (see Table~\ref{tab:bgalzsamples})\footnote{As in Fig.~\ref{fig:bz}, the high-$z$ PRIMUS results are shown without COSMOS, but the upper error bars in the upper panel indicate the stronger luminosity mark correlations when COSMOS is included.  For the color mark correlations (lower panel) the COSMOS field is consistent with the other fields.}. 
$M(r_p)>1$ implies that luminosity and color are correlated with the environment, as expected from the previous results in this paper.  
The fact that the mark correlations are stronger for color indicates that color is a better tracer of environment than is luminosity, consistent with the results of Skibba et al.\ (2013).
In addition, note that the scale dependence in the figure contributes more information about evolving structure formation than the bias evolution (Fig.~\ref{fig:bz}), which is determined by the clustering signal in the linear regime.

For comparison, $z\sim0.05$ SDSS results are shown for the same number density. 
These are consistent with the PRIMUS measurements within the 1-$\sigma$ error bars. 
We also find that the mock catalogs described in Sec.~\ref{sec:mock} (not shown) also yield consistent luminosity and color mark correlations. 

There appears to be no redshift dependence of either the (rank-ordered) luminosity or color mark correlations, within their 1-$\sigma$ error bars. 
This implies that the luminosity- and color-environment correlations, for both small- and large-scale environments ($r_p<$ and $>1$-2$~{\rm Mpc}/h$), do not evolve significantly between $z\sim0$ and $z\sim1$. 
This is consistent with arguments in Cooper et al.\ (2006) and Coil et al.\ (2008), and with recent constraints on the evolution of the environmental dependence of the red or quenched fraction (Woo et al.\ 2013; Kova\v{c} et al.\ 2014). 
Moreover, while the large-scale signal is related to luminosity and color dependent bias, and the luminosity-halo mass and color-halo mass relation of central galaxies, 
the near-constant color mark correlations at small scales suggests that at a given halo mass the red fractions of \textit{satellite} galaxies are not significantly evolving.  
This implies that: some satellites were already red at $z\sim0.8$; red satellites that have merged since then are replaced by newly quenched satellites at lower redshift; and that the quenching efficiency has not evolved much over this time.

\subsection{Implications for Galaxy Evolution}\label{sec:galevol}


Our results on the luminosity and color dependence of galaxy clustering have implications for galaxy evolution, as the pathways galaxies take in the color-magnitude diagram depend on the environment and halo mass (Skibba et al.\ 2009; Schawinski et al.\ 2013). 

Analyses of local galaxy overdensities and mark statistics have shown that, 
although galaxy luminosity and color are independently correlated with the 
environment on small and large scales, the environmental dependence of color is 
stronger than that of luminosity (e.g., Hogg et al.\ 2003; Blanton et al.\ 2005; Skibba et al.\ 2013). 
Our clustering results (in Sec.~\ref{sec:redblue}) are consistent with this, and 
with similar results in the SDSS and DEEP2 (Zehavi et al.\ 2011; Coil et al.\ 2008). 
Therefore, the correlation between color and environment, and the 
environmental dependence of star formation quenching\footnote{Not all red 
sequence galaxies are quenched, as some have obscured star formation (e.g., 
Maller et al.\ 2009; Zhu et al.\ 2011), 
but optical color is nonetheless a good proxy for specific SFR (Mostek et al.\ 2013).}, 
is the crucial correlation to be investigated (see also Kauffmann et al.\ 2004). 

Color dependent clustering is related to the red and blue sequences 
of the color-magnitude distribution, 
such that at a given luminosity, red sequence galaxies tend 
to reside in more massive halos than blue galaxies, and satellite galaxies tend 
to be redder than central ones (Coil et al.\ 2008; Skibba \& Sheth 2009; Loh et al.\ 2010).  
%
Green valley galaxies between the red and blue sequences are often thought to constitute a transitional population in the process of quenching (e.g., Krause et al.\ 2013).  
Central galaxies likely undergo some kind of environmentally dependent transformation as they make this transition, which may be due to major mergers, shock heating, or another mechanism (Skibba et al.\ 2009; Mendez et al.\ 2011; Schawinski et al.\ 2013). 
The build up of the red sequence occurs preferentially in 
relatively dense environments (i.e., in more massive halos), due to the suppression 
of the galaxies' star formation, though there is no preferred halo mass scale, however, 
so it is likely a gradual process (Tinker \& Wetzel 2010; cf., Woo et al.\ 2013; Hartley et al.\ 2013). 

In addition, although the steep correlation functions of red galaxies are consistent with satellite galaxies having a high red fraction, nonetheless a significant fraction of satellites are not yet quenched. 
Moreover, our observed color-dependent clustering is consistent with a model in which the red fraction of satellites is nearly independent of halo mass, indicating that the quenching of satellites' star formation is not a cluster-specific phenomenon, but occurs in groups as well. 
This lends support for `strangulation' (Larson et al.\ 1980) as the primary mechanism for quenching satellites, in which the galaxies' hot gas reservoirs are stripped, thus removing their fuel for future star formation.  
More analysis of stellar mass and SFR dependent clustering (in progress) will shed more light on these issues.

\subsection{Cosmic Variance}\label{sec:var}

It is important to emphasize that our volume-limited PRIMUS samples are 
constructed from seven independent science fields, allowing us to 
reduce, and potentially assess, effects of `cosmic variance' (or more precisely, `sample variance'; Scott et al.\ 1994).  
It is natural to expect field-to-field variations, but as stated in Section~\ref{sec:errors}, rare large structures have been known to affect clustering statistics and their uncertainties (Norberg et al.\ 2011; McBride et al.\ 2011b). 

Depending on a survey's redshift and area, and the bias of objects within it, cosmic variance can be a significant source of uncertainty for measurements of counts in cells and correlation functions (Peebles 1980; Somerville et al.\ 2004). 
For some of our 
correlation function measurements, a significant variation among the fields 
is observed (see Fig.~\ref{fig:wpindfields}).  Of the larger fields (XMM, 
CDFS, and COSMOS), COSMOS is an outlier in many cases, with relatively 
strong clustering, especially for luminous and/or red galaxies at $0.6<z<0.8$ (Figs.~\ref{fig:r0gamma}, \ref{fig:r0redblue}, \ref{fig:bgal}, \ref{fig:r0color}, and \ref{fig:bCcomparison}).  
As noted above, its large volume does not appear to compensate for the effects of cosmic variance (see also Meneux et al.\ 2009). 
In contrast, for example, although the SDSS contains the Sloan Great Wall, which may be the densest structure within the Hubble volume (Sheth \& Diaferio 2011), its large contiguous area in the North Galactic Cap ($\sim7500~\mathrm{deg}^2$; Abazajian et al.\ 2009) does appear to partially compensate for this (though not entirely; see Zehavi et al.\ 2011).

It is interesting that, although we are analyzing the clustering of galaxies that are not highly biased, and in spite of COSMOS's large area (over $1~{\rm deg}^2$)---two factors which would make one expect less significant cosmic variance (Driver \& Robotham 2010; Moster et al.\ 2011)---this field is significantly affected.  
Even the blue galaxies, which do not tend to reside in large structures (except perhaps in their outskirts), nonetheless have a somewhat anomalous clustering signal in COSMOS and an uneven redshift distribution as well.
None of the other fields are so often an outlier; ES1 and DLS are at the low end of some of the clustering measurements, but they are smaller fields and are not outliers.

Therefore, one should use caution when interpreting clustering and large-scale structure analyses in COSMOS: as noted in 
Section~\ref{sec:wp}, it significantly raises the clustering amplitude of 
our composite correlation functions and increases their errors.  
Some authors have nonetheless used COSMOS to study clustering/large-scale structure evolution and 
attempted to constrain models (e.g. Leauthaud et al.\ 2012; Jullo et al.\ 2012; Scoville et al.\ 2013; Tinker et al.\ 2013),  
while others have focused on low-$z$ clustering results and neglected their high-$z$ counterparts, which often have larger statistical and systematic uncertainties, narrower dynamic range, and are more affected by cosmic variance (Neistein et al.\ 2011; Wang et al.\ 2013),  
%
unlike clustering constraints in the local universe. 
It is important to develop such higher-redshift results, however, in order to better constrain analytic and semi-analytic models. The clustering measurements presented here can contribute to this effort.

\section{Summary}\label{sec:conclusions}

In this paper we measure and analyze the luminosity and color dependence of galaxy clustering in the PRIMUS survey, using volume-limited catalogs of over 60,000 galaxies with high-quality redshifts at $0.2<z<1.0$. 
Our analysis includes the study of relatively faint, blue, and low-mass galaxies, with luminosities down to $L\approx0.04L_\ast$ ($M_g=-17$), which have until now not been previously studied by other intermediate-redshift surveys. 

We summarize our main results as follows:
\begin{itemize}
 \item \textbf{luminosity dependent clustering}: 
We find that the clustering strength increases with luminosity at all galaxy separations ($0.1<r_p<30~{\rm Mpc}/h$), with a more rapid increase at $L>L_\ast$.  We also detect a luminosity dependence for blue cloud and red sequence galaxies, which is significantly stronger for the latter.
 
 \item \textbf{color dependent clustering}: 
We find that the clustering amplitude increases significantly with color, with redder galaxies having stronger clustering especially at small scales ($r_p<1~{\rm Mpc}/h$).  We also detect a color dependence at fixed luminosity.  
\textit{Within} the red sequence, we detect a significant color dependence, such that the reddest galaxies are more strongly clustered than their less red counterparts; green valley galaxies cluster approximately intermediately between red and blue galaxies, while we do not detect a significant color dependent clustering within the blue cloud. 

 \item \textbf{clustering evolution}: 
We detect a small amount of evolution in the clustering strength and bias of galaxies selected by luminosity or number density, but our large-scale results are consistent with passive evolution (see Sec.~\ref{sec:bevol}).  The lack of strong evolution implies that the luminosity-halo mass relation does not evolve strongly over $0<z<1$. 

 \item \textbf{halo-model interpretation}: 
We interpret the clustering measurements in terms of `one-halo' and `two-halo' terms, for pairs of galaxies in single or separate halos, with a transition between them at $r_p\sim1$-$2~{\rm Mpc}/h$.  We find that brighter and more massive galaxies tend to be hosted by more massive halos, and tend to have higher fractions of satellite galaxies.
\end{itemize}


This paper is one of a series on galaxy clustering in the PRIMUS survey.  
Bray et al.\ (in prep.) will present complementary results using cross-correlations of PRIMUS galaxies with photometric galaxy samples at small scales.
In addition, work is underway on the clustering properties of X-ray and infrared-selected AGN (Mendez et al., in prep.).  
In future work, we plan to examine the galaxy clustering dependence on stellar mass and star formation, and to further analyze PRIMUS clustering measurements with halo occupation and other models.
These clustering measurements, especially the projected correlation functions, mark correlation functions, and bias, will serve as important constraints on state-of-the-art halo models of galaxy evolution (e.g., Yang et al.\ 2012; Watson \& Conroy 2013) and semi-analytic models (e.g., Q.\ Guo et al.\ 2013; Contreras et al.\ 2013). 

\section*{Acknowledgments}
ALC and RAS acknowledge support from the NSF CAREER award AST-1055081. 
%
We thank Hong Guo, Peder Norberg, and Idit Zehavi for valuable discussions about our clustering results. 
This paper also benefited from discussions with Peter Behroozi, Surhud More, Doug Watson, and others at the ``Galaxies in the Cosmic Web" conference in Chicago. 
We thank the anonymous referee for detailed comments and suggestions for improving the paper. 
We also thank Andrew Hearin and Ashley Ross for helpful comments on a previous draft. 

We acknowledge Scott Burles for his contribution to the PRIMUS project. 
Funding for PRIMUS has been provided by NSF grants AST-0607701, 0908246, 0908442,  0908354, and 
NASA grants 08-ADP08-0019, NNX09AC95G.  This paper includes data gathered with the 6.5~meter Magellan Telescopes located at Las Campanas Observatory, Chile.


\bibliographystyle{apj}

\end{document}